\documentclass[acmsmall]{acmart}

\AtBeginDocument{%
  }

\setcopyright{acmlicensed}
\acmJournal{PACMMOD}
\acmYear{2023} \acmVolume{1} \acmNumber{1}
\acmArticle{111} \acmMonth{5}
\acmPrice{15.00}
\acmDOI{10.1145/3588965}

\usepackage[normalem]{ulem}
\usepackage{subfigure}
\usepackage{bm}
\usepackage{threeparttable}
\usepackage{amsmath}
\usepackage{multirow}
\usepackage{algorithm}  
\usepackage[noend]{algorithmic} 
\usepackage{xcolor}

\newtheorem{definition}{Definition}
\renewcommand{\algorithmicrequire}{\textbf{Input:}}

\begin{document}

\title{Dumpy: A Compact and Adaptive Index for Large Data Series Collections}

\author{Zeyu Wang}
\orcid{0000-0002-0455-0830}
\affiliation{%
  \institution{School of Computer Science, Fudan University}
  \city{Shanghai}
  \country{China}
}
\email{zeyuwang21@m.fudan.edu.cn}

\author{Qitong Wang}
\orcid{0000-0001-6360-3800}
\affiliation{%
  \institution{LIPADE, Universit{\'e} Paris Cit{\'e}}
  \city{Paris}
  \country{France}
}
\email{qitong.wang@etu.u-paris.fr}

\author{Peng Wang}
\orcid{0000-0002-8136-9621}
\authornote{Corresponding author}
\affiliation{%
  \institution{Shanghai Key Laboratory of Data Science, School of Computer Science, Fudan University}
  \city{Shanghai}
  \country{China}
}
\email{pengwang5@fudan.edu.cn}

\author{Themis Palpanas}
\orcid{0000-0002-8031-0265}
\affiliation{%
  \institution{LIPADE, Universit{\'e} Paris Cit{\'e} \& IUF}
  \city{Paris}
  \country{France}
}
\email{themis@mi.parisdescartes.fr}

\author{Wei Wang}
\orcid{0000-0003-0264-788X}
\affiliation{%
  \institution{Shanghai Key Laboratory of Data Science, School of Computer Science, Fudan University}
  \city{Shanghai}
  \country{China}
}
\email{weiwang1@fudan.edu.cn}

\renewcommand{\shortauthors}{Zeyu Wang et al.}

\begin{abstract}
Data series indexes are necessary for managing and analyzing the increasing amounts of data series collections that are nowadays available.
These indexes support both exact and approximate similarity search, with approximate search providing high-quality results within milliseconds, which makes it very attractive for certain modern applications.
Reducing the pre-processing (i.e., index building) time and improving the accuracy of search results are two major challenges.
DSTree and the iSAX index family are state-of-the-art solutions for this problem.
However, DSTree suffers from long index building times, 
while iSAX suffers from low search accuracy.
In this paper, we identify two problems of the iSAX index family that adversely affect the overall performance.
First, we observe the presence of a \emph{proximity-compactness trade-off} related to the index structure design (i.e., the node fanout degree), significantly limiting the efficiency and accuracy of the resulting index.
Second, a skewed data distribution will negatively affect the performance of iSAX.
To overcome these problems, we propose Dumpy, an index that employs a novel multi-ary data structure with an adaptive node splitting algorithm and an efficient building workflow.
Furthermore, we devise Dumpy-Fuzzy as a variant of Dumpy which further improves search accuracy by proper duplication of series.
Experiments with a variety of large, real datasets demonstrate that the Dumpy solutions achieve considerably better efficiency, scalability and search accuracy than its competitors.
This paper was published in SIGMOD'23.
\end{abstract}



\received{July 2022}
\received[revised]{October 2022}
\received[accepted]{November 2022}

\maketitle

\section{Introduction}
\label{sec:intro}
Massive data series collections are now being produced by applications across virtually every scientific and social domain~\cite{talking1,talking2,talking3}, making data series one of the most common data types. 
The problems of managing and analyzing large-volume data series have attracted the research interest of the data management community in the past three decades~\cite{DBLP:conf/sofsem/Palpanas16,DBLP:journals/dagstuhl-reports/BagnallCPZ19}. 
In this context, similarity search is an essential primitive operation, lying at the core of several other high-level algorithms, e.g., classification, clustering, motif discovery and outlier detection~\cite{talking2,DBLP:journals/sigmod/PalpanasB19,app1,app2,app3,DBLP:journals/pvldb/0003WNP22}. 

Similarity search aims to find the nearest neighbors in the dataset, given a query series and a distance measure. The naive solution is to sequentially calculate the distances of all series to the query series. 
However, sequential scan quickly becomes intractable as the dataset size increases.
To facilitate similarity search at scale, a data series index can be used to prune irrelevant data and thus, reduce the effort required to answer the queries. 
Moreover, as researchers pay more attention to data exploration, the importance of approximate similarity search grows rapidly~\cite{hydra1,hydra2}. 
It is widely employed in real-world applications since it can provide high-quality approximate query results within the interactive response time, in the order of milliseconds~\cite{nn-workshop, wang-evaluation,approx1,li-evaluation,elpis}. 
In such applications, approximate query result quality is sufficient to support downstream applications~\cite{knn-prediction}.
Recent examples include 
1) a k-nearest-neighbors (kNN) classifier~\cite{knn-classifier}, whose accuracy converges to the best when kNN mean average precision (MAP) reaches 60\%;
2) an outlier detector~\cite{knn-outlier} that achieves the best ROC-AUC with 50\% MAP;
and 3)
a kNN-based SoftMax approximation technique for large-scale classification, which achieves nearly the same accuracy as the exact SoftMax when kNN recall reaches 80\%~\cite{knn-softmax}.
For these applications, the core requirements for the kNN-index are the query time under the above precision (should be in the order of milliseconds), the index building time, and the scalability to support large datasets.

Although there are dozens of approaches in the literature that can index data series~\cite{hydra2},
only a few of them can robustly support large data series collections, e.g., over 100GB (which is why techniques for approximate search~\cite{hydra2}, as well as progressive search for exact~\cite{pros} and approximate~\cite{DBLP:journals/tvcg/JoSF20} query answering have been studied).
Among them, DSTree~\cite{ds-tree} and the iSAX index family~\cite{c19-isip-Palpanas-isaxfamily} show the best query performance on the approximate search and support exact search at the same time.
Due to the dynamic segmentation technique, DSTree requires a long index building time (over one order of magnitude slower than iSAX) and is hard to optimize.  
On the contrary, benefiting from  fast index building and rich optimizations~\cite{isax2+,ads,DBLP:journals/tkde/YagoubiAMP20,paris+,messi,sing,odyssey}, the iSAX index family has become the most popular data series index in the past decade.
Nonetheless, iSAX still suffers from unsatisfactory approximate search accuracy when visiting a small portion of data (one or several nodes, ensuring millisecond-level delay), e.g., its MAP is less than 10\% when visiting one node while query time exceeds one second when improving MAP to $\geq$50\%~\cite{hydra2}.
In this work, we identify the intrinsic problem of the index structure and building workflow of the iSAX index family and propose our novel solution, Dumpy, to tackle those.

First, we observe that the design of the index structure is an inherent but overlooked problem that significantly limits the performance of the iSAX index family.
Although iSAX~\cite{isax} does not in principle limit the fanout of a node, popular iSAX-family indexes~\cite{isax2.0,ads,DBLP:journals/tkde/YagoubiAMP20,ulisse,paris,messi,sing,odyssey} still adopt a binary structure (except for the first layer that has a full fanout).
When a node contains more series than the leaf size threshold $th$, it selects one SAX segment and splits the node into two child nodes.

However, under this binary structure, the splitting policies being used~\cite{isax, isax2.0} often lead to sub-optimal decisions (cf.~\cite{tardis}, Section~\ref{sec:trade-off}), that hurt the proximity (i.e., similarity) of series inside a node, and finally the quality of approximate query results.

Recently, a full-ary SAX-based index has been proposed to tackle this problem~\cite{tardis}.
A full-ary structure splits a full node on all segments, such that it avoids the problems of focusing on a single segment that leads to sub-optimal splitting decisions.
However, it generates too many nodes (at most $2^w$, $w$ is the total number of segments) in each split, leading to an excessive number of leaf nodes, and hence extremely low leaf node fill factors. 
This leads to an underperforming (disk-resident) index, due to inefficient disk utilization and overwhelming disk accesses. 
Although subtrees in the index can be merged into larger partitions (e.g., 128MB)~\cite{tardis} to reduce random I/Os, it still incurs substantial overhead to store and load its large internal index structure and introduce many application limitations at the same time.

We term the aforementioned problems as the \emph{proximity-compactness trade-off}.
Both proximity and compactness contribute to similarity search since proximity provides closer series to the query and compactness provides more candidate series when visiting a node.

The binary index structure aims at providing compact child nodes, but impairs the accuracy of query results, whereas the full-ary structure splits the node to preserve the proximity of series inside nodes, but fails to provide leaf nodes of high fill factors (i.e., compactness). 
As a result, both structures fail to exploit the proximity-compactness trade-off,
limiting their performance on search accuracy and also building efficiency.

In this work, we break the limits of a single fixed fanout for the iSAX-family indexes and propose an adaptive split strategy that leads to a multi-ary index structure.
Specifically, we design a novel objective function to estimate the qualities of candidate split plans in the aspects of both proximity and compactness.
We use the average variances of data on selected segments to measure the intra-node series proximity and the variance of fill factors of child nodes to measure the compactness.
Moreover, we propose an efficient search algorithm comprised of three speedup techniques to find the optimal split plan according to our quality estimation.

Besides the index structure design, we identify two other problems of the iSAX-index family preventing the best exploitation of the novel adaptive multi-ary index structure.
The first observation is that when the fanout is large (e.g., the first layer in the binary structure and all layers in the full-ary structure), data series are often distributed among the child nodes in a highly imbalanced way, which cannot be entirely avoided, even when we choose the best split plan.
That is, most data series concentrate on only a few nodes while most nodes are slight in size.
It usually leads to a large number of small nodes that impair the performance of the resulting index.
The other problem is that the common iSAX index building workflow splits a node by relying only on the distribution of a tiny portion of data, which actually makes the splitting decisions sub-optimal for the data as a whole. 
For example, iSAX2+ tries to balance two child nodes in splitting according to the first $th+1$ series (i.e., split once it is full), but the final average fill factor is usually less than 20\% as verified in our experiments.

To avoid these two problems, we design a flexible and efficient index-building workflow along with a leaf packing algorithm.
Benefiting from the static segmentation of iSAX, our workflow can collect the global SAX word tables without incurring any additional overhead, and make our adaptive split strategy better fit the whole dataset.
Moreover, our leaf node packing algorithm can pack small sibling leaves without losing the pruning power, contributing to fewer random disk accesses during index building and querying.

In summary, by combining the adaptive split strategy with the new index building workflow, we present our data series indexing solution, Dumpy (named after its short and compact structure).
Dumpy advances the State-Of-The-Art (SOTA) in terms of index building efficiency, approximate search accuracy, and exact search performance, making it a fully-functional and practical solution for extensive data series management and analysis applications.

Moreover, generally as a space-partition-based approach, Dumpy also suffers from a common boundary issue~\cite{nsg,spann}.
That is, the kNN of a query may locate in the adjacent node or subtree and near the partition boundary.
Since we only search one to several nodes, these true neighbors may be missing.
To alleviate this effect, we propose a variant of Dumpy, Dumpy-Fuzzy, which transfers the \emph{hard} partition boundary to a \emph{fuzzy range}, and adopts a duplication strategy in each split to further improve the search accuracy, at the cost of a small overhead on index building and storage.

Our contributions can be summarized as follows.

\noindent(1)
We identify the inherent proximity-compactness trade-off in the structural designs of the current SOTA iSAX-index family, and demonstrate that it limits the quality of approximate query results, as well as the index building efficiency. 

\noindent(2)
We present Dumpy, a novel multi-ary data series index that hits the right balance of the proximity-compactness trade-off by adaptively and intelligently determining the splitting strategy on-the-fly. 

\noindent(3)
We design a powerful and efficient index-building workflow for the iSAX-index family with a novel leaf packing algorithm to handle data skewness and achieve robust performance.

\noindent(4)
We devise Dumpy-Fuzzy to further improve search accuracy by proper data duplication.

\noindent(5)
Our experimental evaluation with a variety of synthetic and real datasets demonstrates that Dumpy and its variants provide consistently faster index building times (4x on average), and higher approximate query accuracy (65\% higher MAP on average) than the SOTA competitors, with query answering times in the order of milliseconds.

\section{Related Work}
\label{sec:related}

\noindent\textbf{[Data series indexes]} 
Dozens of methods have been proposed to index massive data series collections~\cite{hydra1, hydra2}. 
Among these, the SAX-based indexes~\cite{c19-isip-Palpanas-isaxfamily} have gained popularity and achieved SOTA performance. 
Following the initial iSAX~\cite{isax} index, 
iSAX2.0 and iSAX2+~\cite{isax2.0, isax2+} provide faster index building through novel bulk loading and node splitting strategies, 
ADS~\cite{ads} optimizes the combined index building and query answering time, 
ULISSE~\cite{DBLP:journals/vldb/LinardiP20} supports subsequence similarity search, 
SEAnet~\cite{qt} improves query results quality for high-frequency time series using deep learning embeddings, while 
DPiSAX~\cite{dpisax}, Odyssey~\cite{odyssey}, PARIS~\cite{paris+}, MESSI~\cite{DBLP:journals/vldb/PengFP21}, SING~\cite{sing}, and Hercules~\cite{DBLP:journals/pvldb/EchihabiFZPB22} exploit distribution and modern hardware parallelism. 
These indexes all inherit the original binary structure of iSAX, which limits their intra-node series proximity.

ADS~\cite{ads}, as a \emph{query-adaptive} index, builds and materializes only the leaf nodes visited by the examined queries. 
However, in the case of a huge query workload that visits all leaf nodes of the index, ADS becomes the same as an iSAX index, with the same query answering properties. 
On the contrary, as a \emph{data-adaptive} index, Dumpy adapts its structure based on the data collection rather than the queries. 
Therefore, its performance is independent of workloads.
(We omit ADS in the experiments since it is not superior to iSAX2.0 and DSTree~\cite{hydra1}.)

TARDIS~\cite{tardis} first notices the drawbacks of the binary structure and proposes a full-ary structure along with a size-based partitioning strategy to merge different subtrees to be applied in a distributed cluster.
However, TARDIS is only for analyzing a static dataset and the enormous structure decreases the building and query efficiency.
We implement a stand-alone version of TARDIS 
in our experiments.
Coconut~\cite{coconut,DBLP:journals/vldb/KondylakisDZP19} builds a B+-tree after sorting the dataset using the InvSAX representations and gains remarkable performance improvement from sequential I/Os in bulk loading. 
However, the sequential layout on disk will be destroyed by further insertions, and the scan-based exact-search algorithm requires a complete InvSAX table to be kept in memory and the raw dataset in place.
And it seems no easy way to restore the classical tree-based pruning in Coconut.
Hence, we do not include Coconut in our experiments. 

DSTree~\cite{ds-tree} achieves remarkable search accuracy by adopting a highly adaptive summarization EAPCA and increasing the number of segments on the fly.
While the side effect is that DSTree cannot skip costly split operations on raw data series.
Bulk loading algorithms and many other optimizations we mentioned are therefore hard to be applied on DSTree.
As evaluated in our experiments, Dumpy provides higher-quality query results than DSTree even on the static summarization iSAX, with a much faster building time.

\noindent\textbf{[High-dimensional vector indexes]}
According to recent studies~\cite{hydra1, hydra2}, similarity search algorithms for data series and high-dimensional vectors could be employed interchangeably.
Representative algorithms for high-dimensional vector search include proximity graph-based methods~\cite{nsw,hnsw,wang-evaluation}, showing excellent query performance on small datasets, but consuming excessive time and memory to build and store the graph.
Now they are not easy to scale in billion-scale datasets in commodity machines~\cite{diskann, satellite, nsg}.
Product quantization family methods~\cite{pq,opt-pq,inverted,new-pq} achieve better query performance on minute-level near-exact search than data series indexes in advanced research.
However, the building time is still over one magnitude slower than DSTree~\cite{hydra2}.
LSH (Local Sensitive Hashing) family methods~\cite{parsketch,qa-lsh,idec,srs,eilsh}, though providing probabilistic guarantees, have been shown to fall behind data series indexes in terms of time and space~\cite{hydra2}.

\section{Background}
\label{sec:background}

We first provide some definitions necessary for the rest of this paper, and then explain the iSAX summarization and index.

\begin{definition}[Data Series]
A data series $s=\{p_1,p_2,\dots,p_l\}$ is a sequence of points $p_i=(x_i,t_i)$ where each point is associate with a value $x_i$ and a position $t_i$, satisfying that $t_1 < t_2 < \dots < t_l$. $l$ denotes the length of data series $s$.
\end{definition}
In this paper, we assume a data series database $db$ contains numerous data series of equal length $n$.
We use the kNN (k-Nearest Neighbor) query to denote a specific similarity search query with an explicit number of nearest neighbors.
\begin{definition}[$k$NN Query]
Given an integer $k$, a query data series $q$ and a distance measure $dist$, a \textbf{kNN Query} retrieves from the database the set of series $R=\{r_1,r_2,\dots,r_k\}$ such that for any other series $s$ in the database and any $r_i \in R$, $dist(r_i,q) \leq dist(s,q)$. 
\end{definition}
The choice of the distance measure depends on the particular application.
However, Euclidean Distance (ED) is one of the most popular, widely studied and effective similarity measures for large data series collections~\cite{measure}. 
We also support Dynamic Time Warping (DTW)~\cite{dtw} in the meanwhile as the inherent property of the iSAX index family~\cite{messi}.
Besides the exact $k$NN query, the approximate kNN query that accelerates the query processing by checking a small subset of the whole database has attracted intensive interest from researchers. 
The approximate query result, $A=\{a_1,\dots,a_k\}$, is expected to be close to the ground truth result $R$.

\begin{figure}[tb]
\subfigure[SAX of $s$]{
\label{fig:sax3} 
\includegraphics[width=0.35\linewidth]{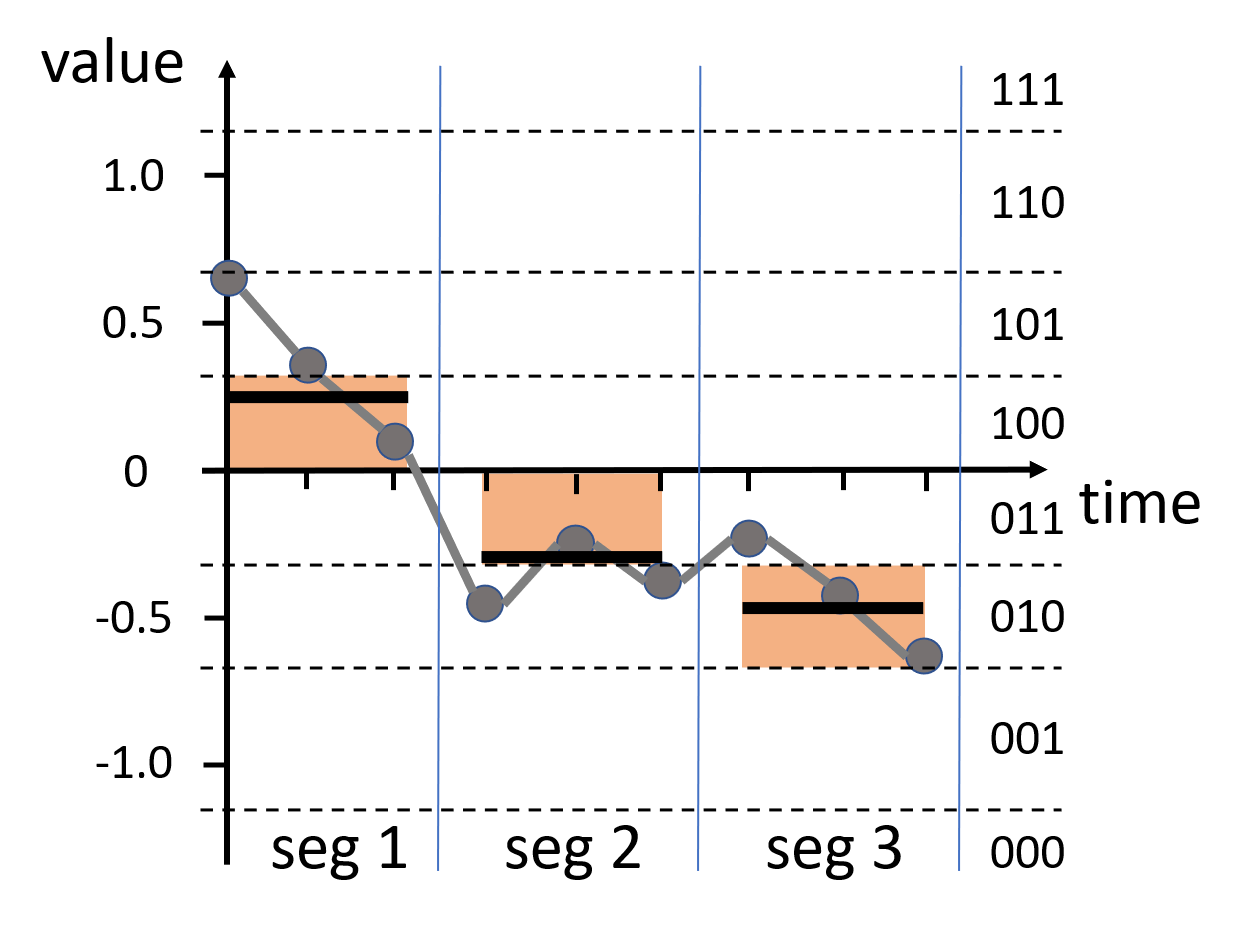}
}
\subfigure[An iSAX word of $s$]{
\label{fig:sax4} 
\includegraphics[width=0.35\linewidth]{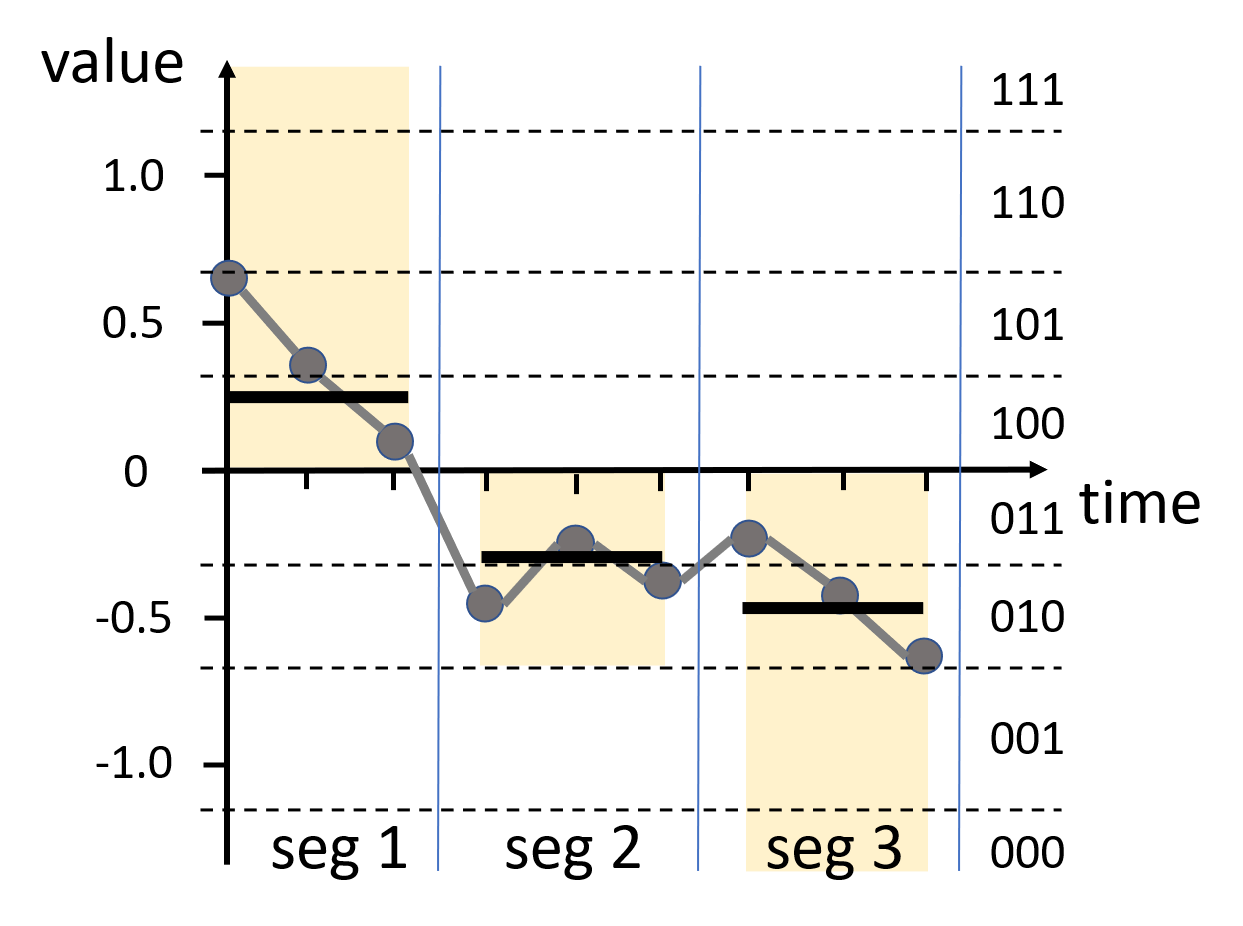}
}

\subfigure[Split on segment 2]{
\label{fig:split2} 
\includegraphics[width=0.25\linewidth]{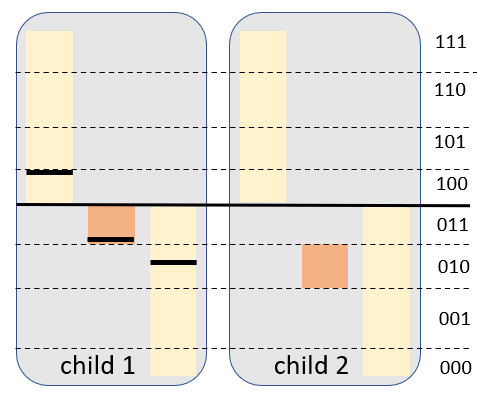}
}
\subfigure[Split on all segments]{
\label{fig:split3} 
\includegraphics[width=0.42\linewidth]{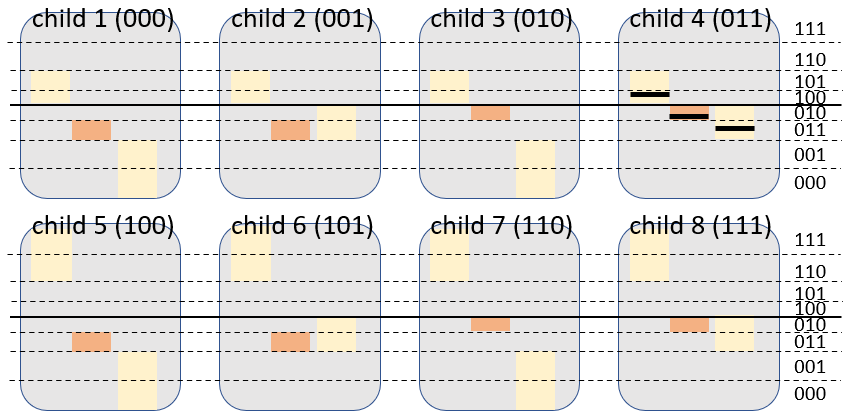}
}
\caption{(a) and (b) are the PAA, SAX and iSAX representation ($w=3$, $b=3$). (c) and (d) are the node splitting for the iSAX-index family in two- and full-ary structures, respectively.}
\label{fig:background} 
\end{figure}

\noindent{\bf [iSAX summarization]}
In this paper, we build Dumpy using the iSAX summarization technique~\cite{isax}.
iSAX is a dynamic prefix of SAX words, and SAX is a symbolization of PAA (Piecewise Aggregate Approximation)~\cite{paa}.
We briefly review these techniques with the example in Figure~\ref{fig:background}.

PAA($s$,$w$) divides data series $s$ into $w$ disjoint equal-length segments, and represents each segment with its mean value. 
Hence, PAA reduces $s$ to a lower-dimensional summarization.
As the black solid line shown in Figure~\ref{fig:sax3}, PAA($s$,3)=[0.28,-0.31,-0.49].

SAX($s$,$w$,$c$) is the representation of PAA by $w$ discrete symbols, drawn from an alphabet of cardinality $c$.
The main idea of SAX is that the real-value space can be split by $c-1$ breakpoints (subject to $N(0,1)$) into $c$ regions, that are labeled by distinct symbols.
For example, when $c$=4 the available symbols (represented in bit-codes) are \{00,01,10,11\}.
SAX assigns symbols to the PAA coefficients on each segment.
In Figure~\ref{fig:sax3}, SAX($s$,3,8)=[100,011,010]. 
The SAX word represents a \emph{region} formed by the value ranges in $w$ segments, drawn in orange background.
iSAX($s$,$w$,$c$) uses variable cardinality ($\leq c$) in each segment.
That is, an iSAX word is a prefix of the corresponding SAX word. 
The iSAX word in Figure~\ref{fig:sax4} is iSAX($s$,3,8)=[1,01,0]\footnote{A special case for the symbol of iSAX word is $*$, at which segment we use only one symbol $*$ ($c=1$) to represent the whole value range.}.
Due to the decreased cardinality of the alphabet, an iSAX word represents a larger range (more coarse-grained) than the corresponding SAX word.

\noindent{\bf [iSAX index family]}
The iSAX index family~\cite{c19-isip-Palpanas-isaxfamily} uses the tree structure to organize data series, which consists of three types of nodes.
The root node representing the whole value space, points to at most $2^w$ child nodes by splitting on all $w$ segments.
Each internal node contains the common iSAX word of all the series in it, and pointers to its child nodes.
Each leaf node, besides the common iSAX word, stores the raw data and SAX words of every series in it.
When the size of a leaf node exceeds the leaf size threshold $th$, the leaf gets transferred into an internal node, and all series in it are split accordingly.
There are two splitting strategies in the iSAX-index family.
One is the binary split (see Figure~\ref{fig:split2}) which splits a node by doubling the cardinality of the iSAX symbol on \emph{one} segment.
These two refined iSAX words representing smaller space on one segment, are assigned to two new leaf nodes.

Considering a node contains the iSAX word in Figure~\ref{fig:sax4} (i.e., [1,01,0]) as an example. The binary strategy splits on the second segment and generates two new leaves with iSAX words [1,010,0] and [1,011,0], respectively.
The classical iSAX index~\cite{isax} adopts the binary split that leads to a binary tree structure below the first layer.
The second strategy is to split on all $w$ segments (i.e., double the cardinality of the iSAX symbol on each segment) and generate at most $2^w$ child nodes (see Figure~\ref{fig:split3}).
This split mode generates a totally full-ary structure, adopted by the recently proposed iSAX-family index, TARDIS~\cite{tardis}.

\section{Proximity-Compactness Trade-off}
\label{sec:trade-off}
We now present the proximity-compactness trade-off, based on the analysis of the binary and full-ary index structures.
More specifically, we claim that neither of them can achieve a high leaf node fill factor (i.e., high compactness) and high intra-node series similarity (i.e., high proximity) simultaneously, which limits the index building efficiency and the approximate query accuracy.

\begin{figure}[tb]
\subfigure[
Skewed region: \emph{a} and \emph{b} are wrongly grouped in node 1-011-0]{
\label{fig:abnormal} 
\includegraphics[width=0.3\linewidth]{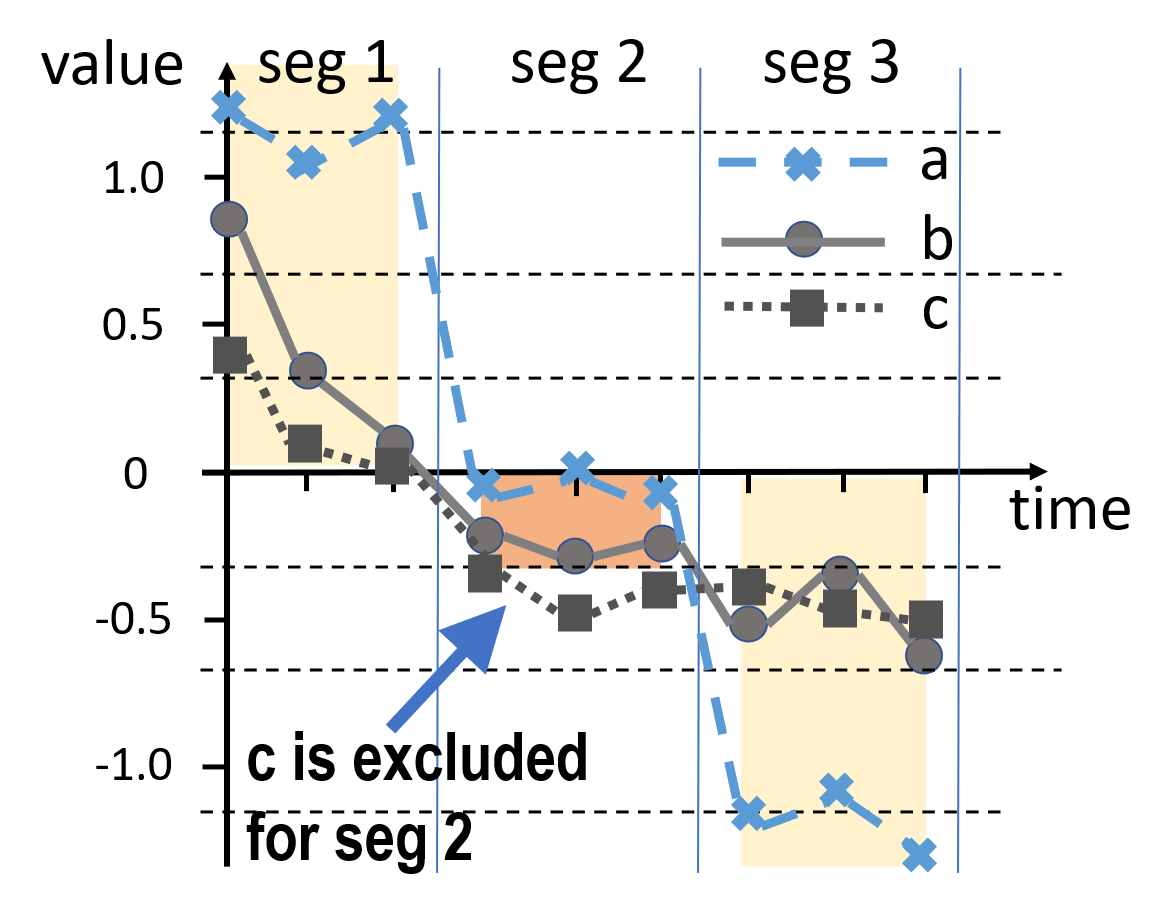}
}
\subfigure[
Even region: \emph{b} and \emph{c} are correctly grouped in node 10-01-01]{
\label{fig:normal}
\includegraphics[width=0.3\linewidth]{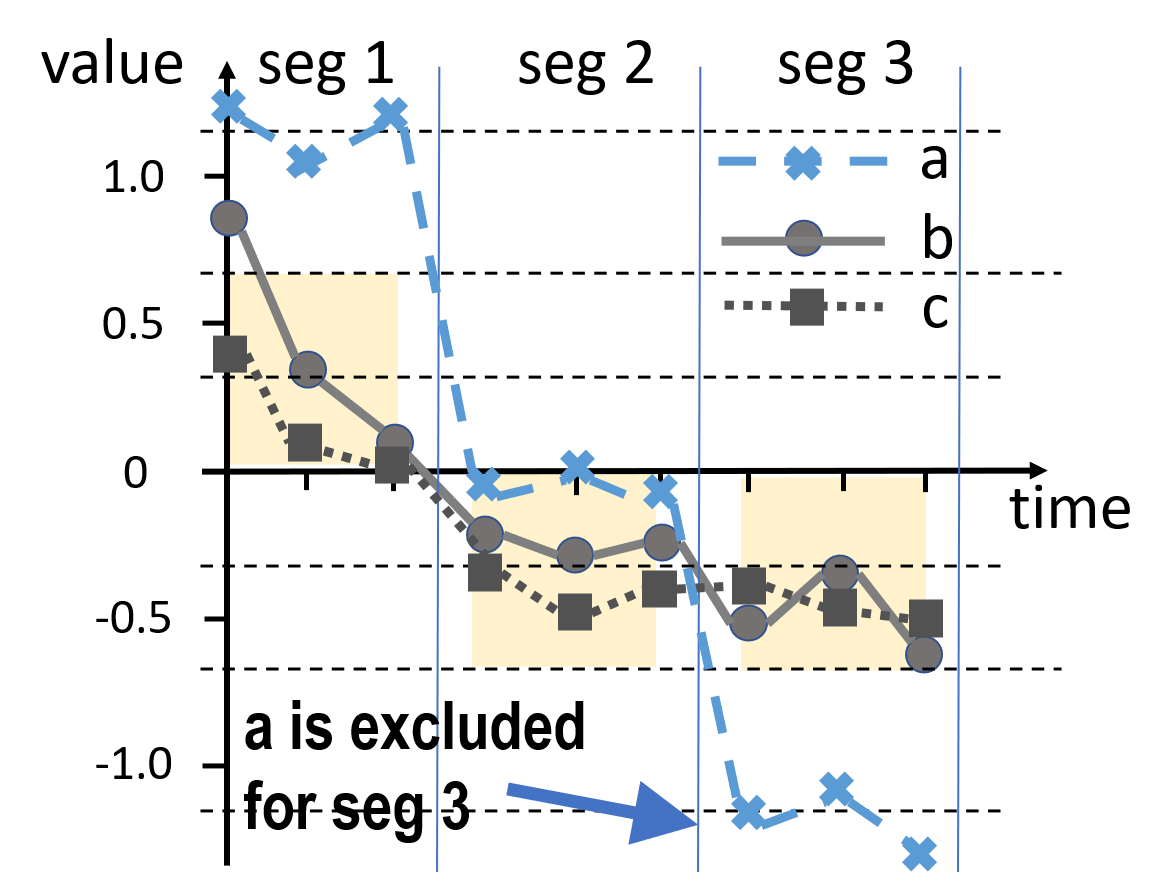}
}
\caption{Illustration of the adverse effect of skewed splits to the intra-node series proximity. Series \emph{b} and \emph{c} are similar to one another, while series \emph{a} is dissimilar to them.}
\label{fig:curse} 
\end{figure}

\noindent\textbf{[Proximity problem of binary structures]}
In a binary structure index like iSAX, the SOTA splitting strategy~\cite{isax2.0} targets to balance the number of series in the two child nodes, by choosing a segment on which the mean value 
is close to the 
breakpoint.
However, this strategy leads to skewed splits: it may split on several specific segments multiple times, leading to an iSAX word 
with several very high-granularity and other very low-granularity segments. 
This situation is depicted in Figure~\ref{fig:abnormal}, where segment 2 has been split three times, 
while the other segments only once. 
Choosing segment 2 may be the best choice for the parent node, yet, this choice is not beneficial for the overall proximity of the series inside the child node.
In our example, the series \emph{b} and \emph{c} are similar overall, but not grouped together due to the slight difference in segment 2, whereas the distant 
series \emph{a} and \emph{b} are grouped into the same node.

Intuitively, this happens because the split decision considers the similarity of the series in an individual segment (segment 2 in Figure~\ref{fig:abnormal}), while proximity is determined by the overall similarity among series across all segments. 
In other words, all segments should be of approximately the same granularity to better reason about similarity (or equivalently, proximity).
On the contrary, a node with a more even subdivision as in Figure~\ref{fig:normal}, will successfully group series \emph{b} and \emph{c} together.
It is important to note that, given a binary fanout, no splitting strategy can provide balanced splits while avoiding the skewness problem.
Thus, binary fanout structures inherently suffer from the proximity problem.

\noindent\textbf{[Compactness problem of full-ary structures]}
Contrary to the binary fanout, a full-ary structure~\cite{tardis} splits a node on all segments.
Hence, it intrinsically avoids the skewness problem by creating a strictly even region. 
However, it quickly generates too many small nodes with low fill factors, severely damaging the index compactness. 
Table~\ref{tab:structure} in our experiments shows the fill factor of a full-ary structure (TARDIS) is below 0.5\% on four public large datasets.
Consequently, the resulting index cannot provide enough candidate series in approximate search, leading to low accuracy when visiting a handful of nodes.
In terms of efficiency, although merging subtrees into larger partitions can significantly reduce random I/Os, storing and loading the enormous structure in a partition file incurs heavy overhead on index building and querying, let alone such dense node packs almost prevent further insertions.

\section{Dumpy}
\label{sec:Dumpy}
In this section, we introduce our solution Dumpy.
Based on a novel adaptive multi-ary structure, Dumpy can hit the right balance of the proximity-compactness trade-off.

In the following, we introduce our basic idea in Section~\ref{sec:motiv} and describe the workflow of Dumpy building in Section~\ref{sec:workflow}.
In Section~\ref{sec:split} we present the adaptive split strategy and the leaf node packing algorithm in Section~\ref{sec:pack}. 
In Section~\ref{sec:search} and ~\ref{sec:update}, we describe the searching and updating algorithm of Dumpy, respectively. 

\begin{figure}[bt]
\subfigure[Random walk]{
\label{fig:dist-rand} 
\includegraphics[width=0.37\linewidth]{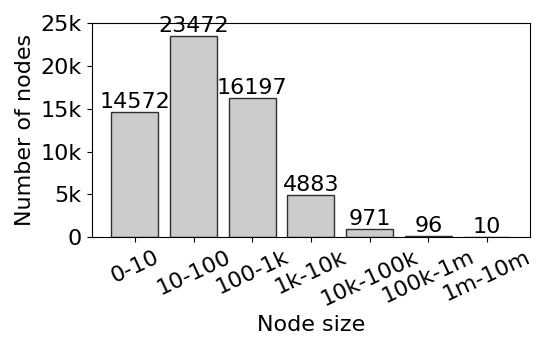}
}
\subfigure[DNA]{
\label{fig:dist-dna} 
\includegraphics[width=0.37\linewidth]{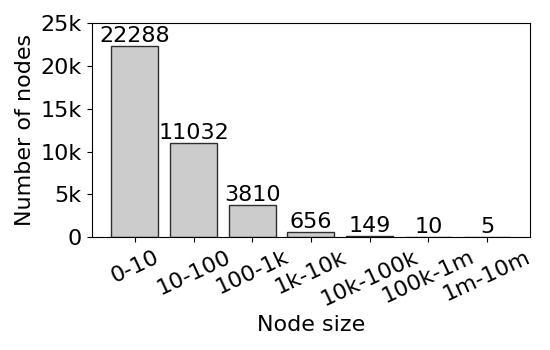}
}
\caption{Node size distribution in the first layer on two 100GB datasets ($w=16$).}
\label{fig:dist-toplayersplits} 
\end{figure}

\subsection{Index Structure and Design Overview}
\label{sec:motiv}

Dumpy organizes data series hierarchically and adopts top-down inserting and splitting as in other SAX-based indexes.
Once a node $N$ is full (its size $c_N$ exceeds the leaf node capacity $th$), Dumpy adaptively selects $\lambda_N$ segments and splits node $N$ on these segments to generate child nodes.
So the fanout of $N$ $\leq 2^{\lambda_N}$.

We demonstrate an example Dumpy tree with $w=4$ segments in Figure~\ref{fig:pipeline}.
The internal node of Dumpy $N$ maintains a list of chosen segments, $csl(N)=[cs_1,cs_2,\dots,cs_{\lambda_N}]$ where $cs_i$ is the \emph{id} of segments (numbered from 1 to $w$), and $csl(N)$ is sorted by the \emph{id} of segments in ascending order. 
When we concatenate the increased bit of each symbol on $csl(N)$, we can get a $\lambda_N$-length bit-code, denoted by \emph{sid} in Dumpy.
Node $N_1$ is an internal node with iSAX word [0,1,1,1].
$csl(N_1)=[1,3,4]$ means 
$\lambda_{N_1}$=3 and we split $N_1$ on segments 1, 3, 4.

In the physical layout, a leaf node corresponds to continuous disk pages storing the raw series and SAX words.
An internal node maintains a hash table to support tree traversal, named \emph{routing table}, mapping \emph{sid} to its corresponding child node.

We now present the intuitions behind our adaptive node splitting algorithm.
To find the best balance between the proximity-compactness trade-off, we design an objective function to evaluate each possible split plan, where we use the variance of data on certain subspace to estimate the proximity of series inside child nodes and use the variance of fill factors of leaf nodes to surrogate the compactness.
Considering the whole search space is $2^w$+1, we first eliminate unpromising plans and then employ the relationship between different split plans to accelerate searching (cr. Section~\ref{sec:split}).

To better facilitate our adaptive splitting algorithm, we propose a new index-building workflow based on the information of all series (cr. Section~\ref{sec:workflow}).
The building workflow of previous SAX-based indexes 
split a node once it is \emph{just} full, i.e., the $th$+1 series arrives.
Considering a node in the index may be mapped by much more series than $th$, the conventional split decisions will lose effect as the first $th+1$ series soon become a small portion of all series falling into this node.

Last but not least, even if supported by the optimal 
adaptive splits,
there might still exist a large number of small leaf nodes.
This is coming from the fact that data series, similar to high-dimensional vectors, are usually unevenly distributed,
generating many different dense and sparse regions~\cite{DBLP:journals/tkde/KornPF01}.
Figure~\ref{fig:dist-toplayersplits} shows the node size distribution in the first layer of iSAX-type indexes.
$>$60\% nodes in Rand and $>$80\% nodes in DNA have $<$100 series while $<$2\% nodes cover $~$80\% series.
To fully avoid this problem, we propose a novel leaf node packing algorithm, to provide high-quality leaf packs by bounding the maximal demotion bits of them (cr. Section~\ref{sec:pack}).

\subsection{Workflow of Dumpy Building}
\label{sec:workflow}

\begin{figure}[tb]
  \includegraphics[width=0.75\linewidth]{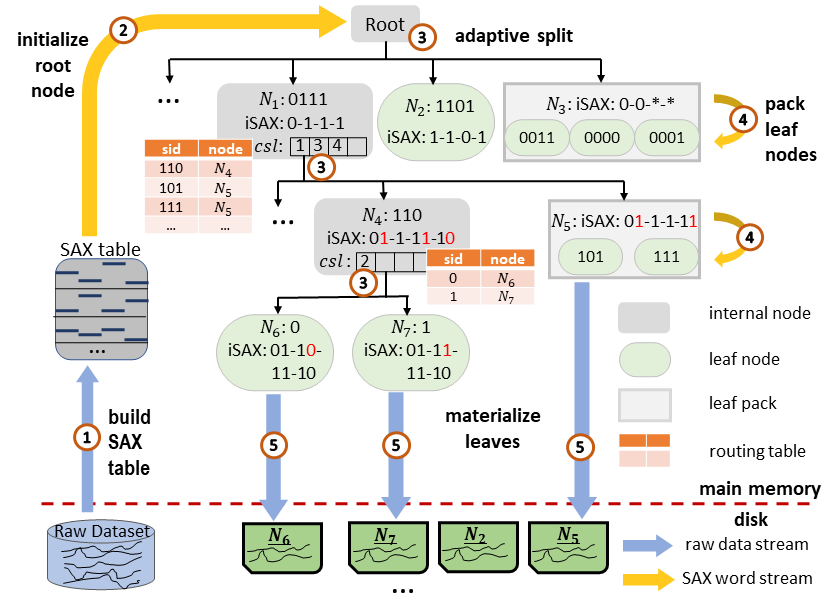}
  \caption{Index structure ($w$=4) and building workflow.
  }
  \label{fig:pipeline}
\end{figure}

The index-building workflow of Dumpy is demonstrated in Figure~\ref{fig:pipeline}.
Dumpy follows the most advanced building framework of the iSAX-family index~\cite{ads-full} but changes two key designs that provide a better index structure and higher building efficiency.
The classical framework is a two-pass procedure.
In the first pass, it reads data series in batch from the raw dataset and computes the SAX words of each series.
Then the SAX words are inserted into the destination leaf node one by one and nodes will be split once it is full.
After the first pass, the index structure is in its final form.
In the second pass, data series are again read in batch, routed to the correct leaf nodes, and written to corresponding files finally.

\noindent{\textbf{[Split nodes using a complete SAX word table]}} 
One key point of the aforementioned framework is to only keep the SAX words in the index (the first pass) and use the SAX words to split nodes, which takes full advantage of the static property of iSAX summarization and significantly reduces disk I/Os.
Dumpy further extends this workflow by separating the SAX words collection and node splitting into two non-overlapping steps, i.e.,
only collecting all SAX words into a SAX table in the first pass and then using
the SAX table to build the index structure before the second pass.
Hence, when splitting a node, 
we know the exact size and the distribution of the series inside it, making our adaptive splitting algorithm take effect actually as we expect.

\noindent{\textbf{[Write to disk after leaf node packing]}}
A large fanout usually generates numerous small nodes before leaf node packing, as shown in Figure~\ref{fig:dist-toplayersplits}.
Considering in the second pass we flush the series of each relevant leaf node in a batch, the number of leaf nodes approximately decides how many random disk writes per batch.
To reduce random writes, before materializing leaves as in the second pass, Dumpy merges sibling small leaf nodes in a proper way to be bigger packs and builds a routing table for the internal node.
Then in the second pass, the series will directly be routed to the leaf pack by the routing table, largely reducing the random disk writes.

\begin{algorithm}[tb]
\caption{Dumpy Construction} 
\label{alg:build}
{\scriptsize
\begin{algorithmic}[1]
\REQUIRE Dataset $db$, leaf node capacity $th$, parameters $r$,$\rho$
\ENSURE Dumpy root $N_r$

\FOR{each series $s$ read in order from $db$}
    \STATE Compute SAX word of $s$, and insert it into SAX table $T_{sax}$ 
\ENDFOR
\STATE Initialize root $N_r$.
\STATE $Split(N_r,T_{sax},th)$
\STATE $NodePacking(N_r, r, \rho, th)$ and label the new leaf node. 
\STATE Initialize an empty hash table as $buf$
\FOR{each series $s$ read in order from $db$}
    \IF{$buf$ is full}
        \STATE flush data in $buf$ to corresponding leaf node files.
    \ENDIF
    \STATE leaf node $N_l$ = $N_i.route2Leaf(T_{sax}[s])$
    \STATE $buf[N_l].insert(s,T_{sax}[s])$
\ENDFOR
\STATE flush $buf$ and clear it.
\STATE free($T_{sax}$)
\RETURN $N_r$
\end{algorithmic}
} 
\end{algorithm}

We now present the complete Dumpy index construction workflow using Algorithm~\ref{alg:build}.
In Stage 1, the SAX table is built from the raw dataset (lines 1-2), and in Stage 2 the root node is initialized (line 3). 
In Stage 3, the adaptive split is first executed on the root node, and then recursively on all internal nodes whose size exceeds $th$ (line 4 and Algorithm~\ref{alg:split}).
In Stage 4, we traverse the index tree and pack leaf nodes under the same parent (line 5 and Algorithm~\ref{alg:pack}).
Now we finish the construction of index structure.
In stage 5 we materialize leaves (line 6-12).
Line 6 prepares a buffer for leaf nodes. 
Then the raw series are read from the disk, routed to leaf nodes, and cached in the buffer along with its SAX words (lines 7, 10-11).
Once the buffer is full, it flushes data to the corresponding file of each node (lines 8-9).

\subsection{Adaptive Node Splitting} 
\label{sec:split}
We now present our adaptive strategy of determining fanouts and splits (on which segments) based on the SAX words of all relevant series.
Our strategy is to select the best split plan based on a novel objective function, which considers the proximity of series inside child nodes and the compactness of child nodes at the same time.
Since the number of all possible split plans is as large as $2^w-1$, we also propose an efficient search algorithm by restricting the candidate space and reusing the shared information.

\begin{figure}[bt]
\subfigure[Balanced split with large variance]{
\label{fig:heuristic-1} 
\includegraphics[width=0.37\linewidth]{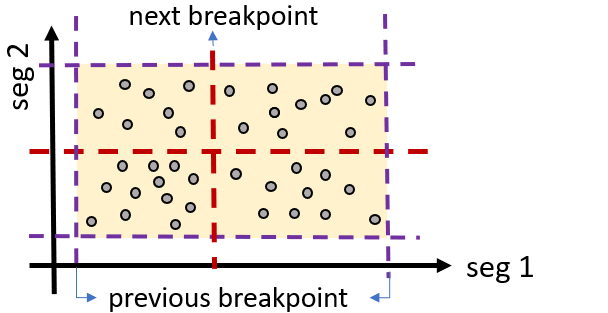}
}
\subfigure[Imbalanced split with large variance]{
\label{fig:heuristic-2} 
\includegraphics[width=0.37\linewidth]{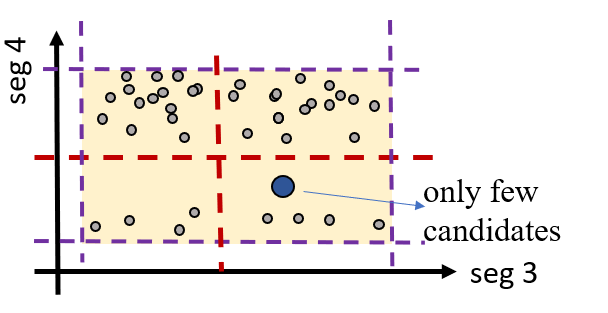}
}

\subfigure[Balanced split with small variance]{
\label{fig:heuristic-3} 
\includegraphics[width=0.37\linewidth]{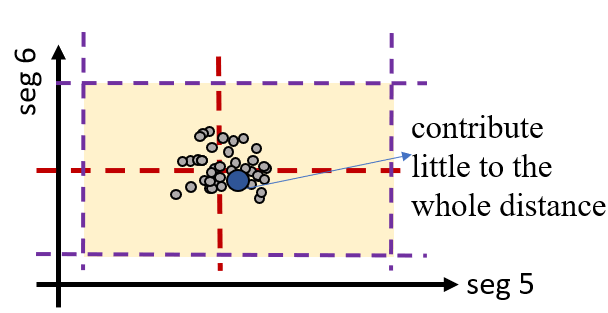}
}
\subfigure[Imbalanced split with small variance]{
\label{fig:heuristic-4} 
\includegraphics[width=0.37\linewidth]{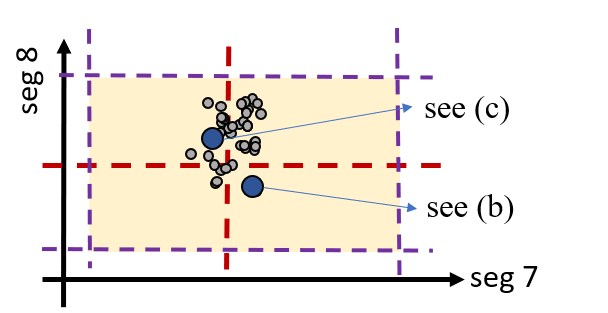}
}
\caption{Four typical scenarios while choosing the segments to split.
  The grey points represent the same set of data series under four different split plans. 
  The blue points are the query series.}
\label{fig:heuristic} 
\end{figure}

\subsubsection{Objective Function}
Our objective function targets to achieve the best balance between the trade-off of proximity and compactness.
We measure the proximity based on the average variances of data on candidate segments.
And to measure compactness, we consider both the variance of fill factors of child nodes and the ratio of overflowed nodes to pursue a balanced split and avoid the bias for small or large fanouts.
We now give the formal definition of the proposed objective function, and then introduce the design principles in detail.

Given a node $N$ containing $c_N$ series $\mathcal{X}_{N} = \{\boldsymbol{x^1},\boldsymbol{x^2},\dots,\boldsymbol{x^{c_N}}\}$ where $\boldsymbol{x^i}=SAX(s^i,w,c)$ is the SAX word of series $s^i$ and $x^i_j$ is $j$-th symbol of $\boldsymbol{x^i}$, and a split plan $csl(N)=[cs_1,cs_2,\dots,cs_{|csl(N)|}]$, we first project each series of $\mathcal{X}_{N}$ onto the segments of $csl(N)$ and get $\mathcal{X'}_{N}$, that is, $x'^{i}_{j}=x^{i}_{cs_j}$ for any $i$ and $1\leq j\leq|csl(N)|$.
Then our objective function is as follows:
\begin{equation}
\label{equ:score}
\mathop{max}\limits_{csl(N)}\ e^{\sqrt{\frac{1}{|csl(N)|}Var(\mathcal{X'}_{N})}} + \alpha*e^{-(1+o)\sigma_{\boldsymbol{F}}}
\end{equation}
\noindent where $e$ is the Euler's number, the variance of projected data is defined as $Var(\mathcal{X'}_{N})=\frac{1}{c_N}\sum_{i=1}^{c_N}\Vert\boldsymbol{{x'}^i}-\boldsymbol{\mu}\Vert^2$ and $\boldsymbol{\mu}$ is a vector of mean values of data on each chosen segments\footnote{We use the midpoint of the range represented by the SAX symbol to calculate the mean value and other statistics.}, $o\in[0,1]$ is the ratio of overflowed child nodes (size $>th$), $\sigma_{\boldsymbol{F}}$ is the standard deviation of the fill factors of child nodes, and $\alpha$ is a weight factor to balance the influence of these two measurements.

The first term estimates the proximity of a split plan.
It evaluates the average variance of relevant data series on the projected SAX space, which is equivalent to the average distance of all the data series to their centroid, i.e., the mean vector $\boldsymbol{\mu}$.
The variance is an indicator of data informativeness on certain dimension~\cite{opt-pq,new-pq,kd-tree}, considering large variances usually mean large information entropy~\cite{entropy}.
We also explain this in Figure~\ref{fig:heuristic}.
Since different plans may choose different numbers of segments, we divide the variance by the number of chosen segments to make the evaluation fair.

The second term is to evaluate the compactness of a split plan.
The standard deviation of fill factors of child nodes prevents extremely imbalanced splits and avoids the severe data skewness like Figure~\ref{fig:dist-toplayersplits}: the value will be very large in this case.
Informally, the vector of fill factors is defined as $\boldsymbol{F} = (F_1,F_2,\dots,F_{2^{|csl(N)|}})$ where $F_i = c_{N_i} / th$ and $N_i$ is the $i$-th child node. 
However, it shows bias for small fanout, which generates fewer but larger child nodes and leads to an unnecessary deep tree.
To resolve this problem, we add a penalty term (1+$o$) that uses the ratio of overflowed child nodes to avoid the bias for plans of small fanout.

We further illustrate the heuristic behind the objective function in Figure~\ref{fig:heuristic}, where we present four different split plans for the same group of data series.
Without loss of generality, we assume all plans choose two segments, i.e., $|csl(N)|=2$.
Then we project the relevant data on a two-dimensional SAX space (different spaces for different plans).
The plan chosen by our objective function is Figure~\ref{fig:heuristic-1}, which has a large variance and balanced child nodes.
The plan in Figure~\ref{fig:heuristic-2} is inferior to Figure~\ref{fig:heuristic-1} due to the imbalance of child nodes, which leads to lower search accuracy for a lack of sufficient candidate series.
The problem of the plan in Figure~\ref{fig:heuristic}(c) is its small variance, which means that series are rather close to each other in these dimensions, on which the distance between series contributes little to the whole distance that is accumulated by all $w$ dimensions.
That is, the distance of query and series in this subspace (small or large) is not informative for their actual distance in the whole space.
Specifically, close series in this subspace may have a large distance actually.
Therefore, the proximity of the series inside a node is crucially weakened.
Figure~\ref{fig:heuristic-4} shows a worst split plan candidate, and our objective function can easily eliminate it.

\subsubsection{Find the Optimal Split Plan}
A naive solution to finding the optimal split plan under our objective function is to iterate and evaluate all $2^w$-1 split plans.
To evaluate the objective function of each plan, it needs at least four passes of all series, rendering the CPU calculation a bottleneck in index building.
To reduce the complexity, we propose a novel searching algorithm composed of three practical speedup techniques, that are, pre-computing the variance for each segment, restricting the search space by a user-defined fill-factor range, and hierarchically computing the sizes of  child nodes.
We first present each technique in detail, then wrap them up in the complete node splitting algorithm 
in Algorithm~\ref{alg:split}.

\noindent{\textbf{[Pre-compute variance]}}
We find that in the first term of the objective function, $Var(\mathcal{X'}_{N})$ can be computed by linearly accumulating the variance of data on each segment.
\begin{equation}
Var(\mathcal{X'}_{N})=\sum_{cs\in csl(N)}Var(\Pi_{cs}(\mathcal{X}_N))
\end{equation}
\noindent where $\Pi_{cs}(\mathcal{X}_N)$ indicates the projection of $\mathcal{X}_N$ onto segment $cs$.

\proof{\em 
$$\sum_{cs\in csl(N)}Var(\Pi_{cs}(\mathcal{X}_N)) = \frac{1}{c_N}\sum_{cs}\sum_{i=1}^{c_N}(x^i_{cs}-\mu_{cs})^2$$
$$=\frac{1}{c_N}\sum_{i=1}^{c_N}\sum_{cs}(x^i_{cs}-\mu_{cs})^2=\frac{1}{c_N}\sum_{i=1}^{c_N}\Vert\boldsymbol{{x'}^i}-\boldsymbol{\mu}\Vert^2=Var(\mathcal{X'}_{N})$$
}

Hence, we can pre-compute the variance of data series on each segment when we start to split a node.
When evaluating a specific plan, we simply fetch the corresponding segments' variances and sum them up with constant complexity.

\noindent{\textbf{[Restrict the search space]}}
We now consider pruning impractical plans based on simple heuristics.
Taking two extreme cases as examples: a plan that splits a node of size $th$+1 on $w$ segments will generate excessively small nodes, and a plan that splits a million-sized node on one segment will generate two huge nodes of size far exceeding $th$.
Hence, it is natural to restrict the average fill factor of child nodes to be in a reasonable range and avoid the particular evaluation.
We introduce a pair of parameters $F_l$, $F_r$ to bound the average fill factor of child nodes.
Then the range of the number of chosen segments $|csl(N)|$ can be deduced as 
\begin{equation}
\label{equ:fill}
\max(1, \log{\frac{c_N}{F_r*th}}) \leq |csl(N)| \leq \min(w, \log{\frac{c_N}{F_l*th}})
\end{equation}
In practice, we empirically set $F_l=50\%$ and $F_r=300\%$, which generally achieves 16x speedup and 99\% accuracy on average.
The acceleration is especially remarkable on relatively small nodes, whose pruned search space is very small.

\begin{figure}[tb]
  \includegraphics[width=0.75\linewidth]{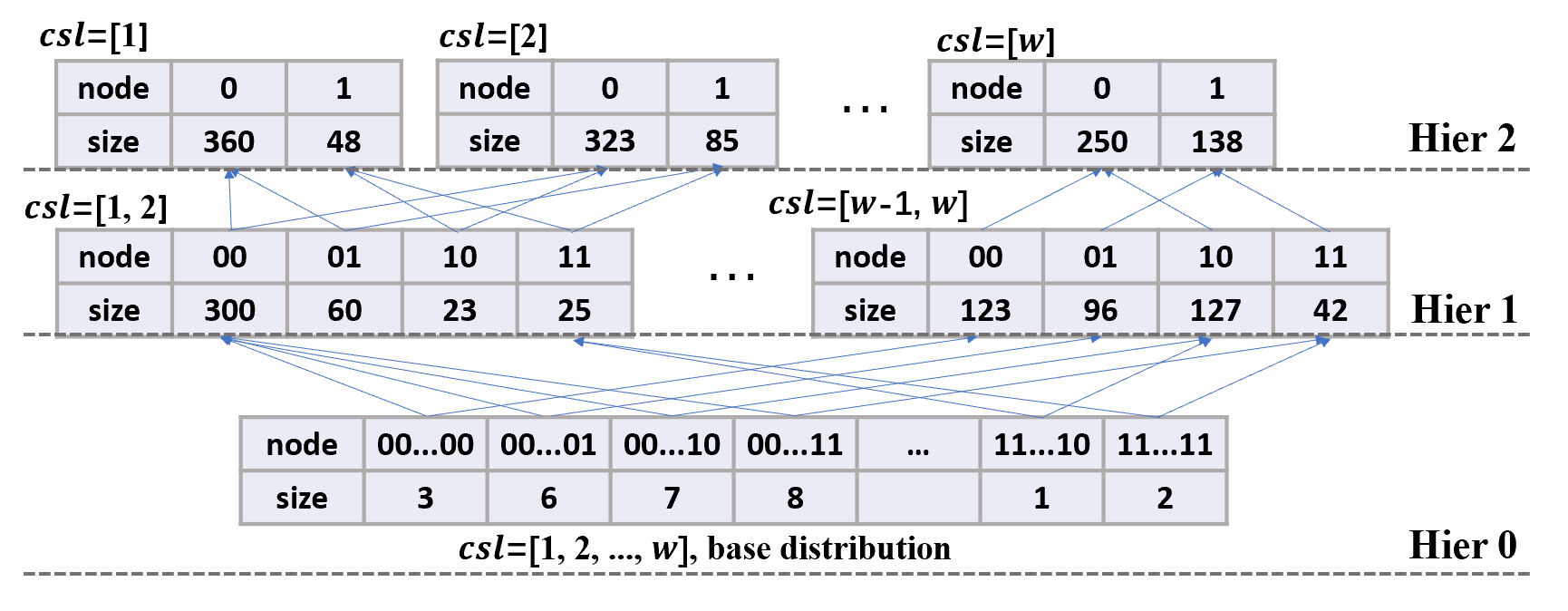}
  \caption{Hierarchically compute child node sizes for 
  each candidate split plan.}
  \label{fig:hier}
\end{figure}

\noindent{\textbf{[Hierarchically compute sizes of child nodes]}}
So far for each plan, we still need to iterate all the data series to get the sizes of child nodes.
We observe that if a split plan $csl^i(N)$ is a subset of another plan $csl^j(N)$, then the size distribution of child nodes of plan $csl^j(N)$ can be reused to calculate the distribution of plan $csl^i(N)$.
Supposing $csl^i(N)=[1]$ and $csl^j(N)=[1,2]$ (the upper left part in Figure~\ref{fig:hier}), and the sizes of four child nodes under $csl^j(N)$ are $c_{N_{00}}=300$, $c_{N_{01}}=60$, $c_{N_{10}}=23$, $c_{N_{11}}=25$,
(nodes are represented by their $sid$s), 
then the sizes of two child nodes under $csl^i(N)$ can be computed as $c_{N_0} = 300+60=360$ and $c_{N_1}=23+25=48$.
Since the whole $w$ segments are a superset of split plans, 
we first compute child node sizes for $w$ segments as a base distribution in each split
and then traverse other plans in a depth-first manner, starting from the plan with the largest fanout to the smallest.
Hence, we can reuse the size distribution we have gained
in a hierarchical way and avoids 
traversing all the series for each plan.

We summarize our splitting strategy in Algorithm~\ref{alg:split}.
The backbone is in lines 4-13 and 19-42 while lines 14-16 split data according to the split plan and lines 17-18 recursively split the overflowed child nodes.
We first prepare the data variance on each segment (lines 4-6) and the base distribution (lines 7-10), and then determine the range of the fanout (line 11).
Then we evaluate each split plan in a hierarchical way (lines 19-42) with a hash set to avoid duplicated evaluation.
Lines 21-26 iterate each combination from the base plan as the current split plan and compute its distribution.
After that, the objective function is evaluated to update the best-so-far answer.
Finally, the sub-plans of the current plan are evaluated similarly.

\begin{algorithm}[tb]
\caption{Split} 
\label{alg:split}
{\scriptsize
\begin{algorithmic}[1]
\REQUIRE node $N$, SAX table $T_{sax}$, leaf node capacity $th$
\IF{$N$ is root node}
    \STATE $csl(N)$ = $[1,2,\dots,w]$
\ELSE
\STATE segVar = []
\FOR{$i$ = 1 to $w$}
    \STATE segVar.push(variance of data in segment $i$)
\ENDFOR
\STATE Initialize $baseDist$ as a list of size $2^w$: [0,0,...,0].
\FOR{each data series $s$ in $N$}
    \STATE $sid$, $\_$ = $promoteiSAX(iSAX(N), T_{sax}[s], [1,2,...,w])$
    \STATE $baseDist[sid]$++
\ENDFOR
\STATE Calculate $\lambda_{max}$,$\lambda_{min}$ according to Equation~\ref{equ:fill}
\STATE Initialize $visit$ as an empty set
\STATE $csl(N)$ = calcDist($baseDist$, $[1,2,...,w]$, $\lambda_{max}$, $\lambda_{min}$, $visit$, $segVar$, null)
\ENDIF
\FOR{each data series $s$ in $N$}
    \STATE $sid$, $isax_{new}$ = $promoteiSAX(iSAX(N), T_{sax}[s], csl(N))$
    \STATE Insert $s$ to $N.children[sid]$ 
    \STATE Label $N.children[sid]$ with $isax_{new}$
\ENDFOR
\FOR{each of $N$'s child node $N_i$ satisfying $c_{N_i}$ $>th$}
    \STATE $Split(N_i, T_{sax},th)$
\ENDFOR
\hrulefill
\renewcommand{\algorithmicrequire}{\textbf{Function}}
\REQUIRE calcDist($baseDist$, $basePlan$, $\lambda_{cur}$, $\lambda_{min}$, $visit$, $segVar$,$bsf$)
\IF{$\lambda_{cur}$ < $\lambda_{min}$}
    \RETURN $bsf$
\ENDIF
\FOR{each combination of cardinality $\lambda_{cur}$ from $basePlan$ as $csl^{cur}$}
    \IF{$visit$ does not contain($csl^{cur}$)}
        \STATE Initialize $Dist$ as a list of size $2^{\lambda_{cur}}$: [0,0,...,0]
        \FOR{each $item$ in $baseDist$}
            \STATE $sid_{new}$ = extract bits in $csl^{cur}$ from $item.sid$
            \STATE $Dist[sid_{new}]$+=$item.size$
        \ENDFOR
        \STATE Compute $score$ using Equation.~\ref{equ:score}
        \IF{score > $bsf$.score}
            \STATE $bsf$ = $csl^{cur}$
        \ENDIF
        \STATE calcDist($Dist$,$csl^{cur}$, $\lambda_{cur}$-1, $\lambda_{min}$, $visit$, $segVar$, $bsf$)
    \STATE $visit$.insert($csl^{cur}$)
    \ENDIF
\ENDFOR
\RETURN $bsf$

\hrulefill
\renewcommand{\algorithmicrequire}{\textbf{Function}}
\REQUIRE promoteiSAX(iSAX word $isax$, SAX word $sax$, chosen segments list $csl$)
\STATE $sid$ = 0
\STATE $isax_{new}$ = $isax$
\FOR{each segment $seg$ in $csl$}
    \STATE $nb$ = $len(isax[seg])$
    \STATE $sid$ = $sid << 1$ + the $(nb+1)$-th bit of $sax[seg]$
    \STATE $isax_{new}[seg]$ = first $nb+1$ bits of $sax[seg]$
\ENDFOR
\RETURN $sid$, $isax_{new}$

\end{algorithmic}
} 
\end{algorithm}

\subsection{Leaf Node Packing}
\label{sec:pack}
As mentioned in Section~\ref{sec:motiv}, data series are usually distributed unevenly in child nodes when the fanout is large, leading to a large number of small leaf nodes and hence, degeneration of the indexing performance.
TARDIS~\cite{tardis} adopts a node-size-based strategy to merge subtrees into a partition, while Coconut~\cite{coconut} proposes a sorting technique for iSAX words, both of which crucially weaken or even abandon the \emph{pruning power} of resulting index. 
To tackle this problem, we propose a simple yet effective algorithm to pack small leaf nodes without losing the pruning ability of packed nodes.

The intuition is to minimize the value range in the SAX space occupied by the packed nodes, i.e., make them have the tightest iSAX representation. 
Tighter iSAX representation directly translates to higher pruning power.
For example, merging two nodes with $sid$ $0010$ and $0100$ is better than merging $0010$ and $0101$, since we demote two bits and get $0**0$ instead of demoting three bits and getting the coarser $0***$.
We define the demotion bits as the different bits between the $sid$s of two or more nodes considered to be merged into the same pack.
In our node packing algorithm, we limit the number of demotion bits to be smaller than $\rho \lambda$, where $\rho$ is a user-defined parameter trading off pack quality and fill factor.
Specifically, given a list of packs and a small node $N$ to be packed, we decide $N$'s belonging by the demotion cost, which is defined as the increased number of demotion bits of the pack if we add $N$ into it.
A leaf node pack forbids any node's insertion request if it will make the pack demote more than $\rho\lambda$ bits or overflow (size > $th$).
Finally, if  no existing pack can satisfy the requirements to insert $N$,
we will create a new pack and insert $N$ into it.

The details of this algorithm are shown in Algorithm~\ref{alg:pack}.
We first sum the sizes of all small leaf nodes (lines 1-5) and randomly initialize the pack list with $\lfloor sum\_size/th \rfloor$ leaf nodes, which is the least number to hold the small nodes (line 6).
Then we iterate all the other small nodes and select or create a pack for them (lines 7-20).
For node $N$, we check each of the existing packs in the pack list and select the one with the least demotion cost (lines 14-16, 19-20).
When the pack demotes more than $\rho*\lambda_{N_p}$ bits, we simply give up this choice (lines 11-12).
Besides, we also ensure the size of all the packs will not exceed $th$ leading to new splits (lines 11-12).
When there is no qualified pack, we will create a new pack initialized with node $N$ (lines 17-18).
Finally, we update the routing table of $N_p$.

The definition of small nodes can be defined by use cases. 
For a static dataset, the small node threshold $r$ can be set to 1 to improve performance, while in a dynamic dataset, $r$ can be set dynamically according to the historical updating frequency.
That is, small $r$ for intensive updating use cases and vice versa.

\begin{algorithm}[tb]
\caption{Node Packing} 
\label{alg:pack}
{\scriptsize
\begin{algorithmic}[1]
\REQUIRE parent node $N_p$, small node threshold $r$, demotion ratio $\rho$, leaf capacity $th$
\STATE $nodes$ = [], $sum\_size$ = 0
\FOR{each of $N_p$'s child node $N_l$ that is not splitted}
    \IF{$c_{N_l}$ < $r*th$}
        \STATE $nodes$.push($N_l$)
        \STATE $sum\_size$ += $c_{N_l}$
    \ENDIF
\ENDFOR
\STATE Initialize pack list $pl$ with $\lfloor sum\_size/th \rfloor$ leaf nodes 
\FOR{each node $N$ in $nodes$}
    \STATE $bsf$ = (null, $\lambda_{N_p}$)
    \FOR{each pack $p$ in $pl$}
        \STATE $p'$ = $p.insert(N)$
        \IF{$c_p'$>$th$ or number of demotion bits of $p'$ > $\rho*\lambda_{N_p}$}
            \STATE continue
        \ELSE
            \STATE $cost$ = increased number of demotion bits in $p'$
            \IF{$cost$ < $bsf$.cost}
                \STATE $bsf$ = ($p$, $cost$)
            \ENDIF
        \ENDIF
    \ENDFOR
    \IF{$bsf$.pack == null}
        \STATE $pl$.push(new pack initialized with $N$)
    \ELSE
        \STATE $bsf$.pack.insert($N$)
    \ENDIF
\ENDFOR
\STATE update $N_p$'s routing table
\FOR{each of $N_p$'s child node $N_i$ satisfying $c_{N_i}>th$ }
    \STATE $NodePacking(N_i, r, \rho, th)$
\ENDFOR
\end{algorithmic}
} 
\end{algorithm}

\subsection{Search Algorithm}
\label{sec:search}
Dumpy supports 
two styles of query answering algorithms.
The first style follows the classical pruning-based search algorithm~\cite{hydra2}.
As a SAX-based index,
Dumpy can conduct an efficient search (including the exact, $\delta$-$\epsilon$-approximate search and etc.~\cite{isax, hydra2}) by pruning irrelevant leaf nodes using lower-bounding distances of iSAX words~\cite{isax, hydra2}.
Besides that, Dumpy also supports traditional approximate search, i.e., querying one target leaf node.
Moreover, we extend it to allow searching more nodes, called \emph{extended approximate search}, to improve query answer quality while maintaining response time in milliseconds.
We limit the search range of extended approximate search in the smallest subtree of the target leaf node to reduce the complexity and avoid traversing the whole tree and evaluating the nodes one by one as in the bound-based search style.
Benefiting from Dumpy's multi-ary structure and fill factor, it brings prominent improvement in search accuracy.

As shown in Algorithm~\ref{alg:search}, the input includes an additional parameter that restricts the visited node number.
The search process starts from the root node and ends with a node that has fewer nodes than the input or an empty node (lines 1-4).
Then we sort the sibling of the ending node according to their lower bound distance (as described in ~\cite{isax} for ED and ~\cite{messi} for DTW) in ascending order (line 5).
Finally, we iterate the sorted sibling nodes or subtrees until reaching the maximal visited node number and return $k$NN (lines 6-9, the concrete search procedure is omitted since it is the same as in other search algorithms).

\begin{algorithm}[tb]
\caption{Extended Approximate Search} 
\label{alg:search}
{\scriptsize
\begin{algorithmic}[1]
\REQUIRE root node $N_r$, node number $nbr$, query series $q$
\STATE node $N$ = $N_r$
\WHILE{$N$!= null and $N$.leafNbr > $nbr$}
    \STATE $sid$ = $promoteiSAX(iSAX(N),SAX(q),csl(N))$
    \STATE $N$ = $N.routingtable[sid]$
\ENDWHILE
\STATE sort $N$'s siblings according to lower bound distance into list $l$
\WHILE{number of searched nodes < $nbr$}
    \STATE $N_c$ = pop the head node of $l$
    \STATE fetch all nodes rooted at $N_c$ and search the series inside
\ENDWHILE
\RETURN $k$NN among the visited series
\end{algorithmic}
} 
\end{algorithm}

\subsection{Updates}
\label{sec:update}
As a fully functional index, Dumpy also supports updates (insertion and deletion) besides bulk loading.
A major difference is that Dumpy no longer collects the information of all series beforehand
on a dynamic dataset.
Though, since all the SAX words are stored in leaves along with the raw series (as the same with iSAX-index family~\cite{isax}), we can re-organize the index structure when an internal node's fanout and size do not satisfy the constraint in Equation~\ref{equ:fill}.

Specifically, we read all the SAX words in the leaves rooted at this node and follow the same workflow in Algorithm~\ref{alg:build}.
Note that this process can be executed in a background thread without blocking the front-end service.
During this period, the query series of updates will come into both the old structure and the new but unfinished one.
Once the background work is finished, we replace the old subtree with the new subtree, and free all the space occupied by the old one.
Another difference is when the query series falls into a full pack.
Then we can simply extract the target leaf node in the pack and redo the node packing for the siblings after a large number of such extractions.
These operations are very fast since these small nodes usually cover a small number of series.

The deletion is almost the same as the iSAX-index family~\cite{isax,ads}.
In particular, we mark the deleted data series in the corresponding leaf via a bit-vector and further insertions can exploit the space occupied by the deleted series while queries ignore these entries.
When a node is empty, we free the occupied space.
The only difference for Dumpy is to update the routing table.

\subsection{Complexity Analysis}
\label{sec:theory}
In this section, we first analyze the time complexity of index building and querying, and then analyze the space complexity.

\noindent{\textbf{[Time complexity]}}
As a disk-based index, the time cost of Dumpy depends on both in-core complexity and disk accesses.
In the following, we first discuss the theoretical time cost in index building and then querying.

The complexity of Dumpy index building could be summed over sub-modules. 
In the adaptive split algorithm, let node $N$ is to be split, and $|csl(N)|$ is between $\lambda_{min}$ and $\lambda_{max}$ according to Equation~\ref{equ:fill}. 
Then the cost of computing the variance of each segment, the base distribution and the routing target is $O(4wc_N)$. 
The number of calculations of the $calcDist$ function is $\binom{w}{\lambda_{max}}*2^{w} + \sum_{i=\lambda_{min}}^{\lambda_{max}-1}\binom{w}{i}2^{\lambda_{i+1}}$, where the first term corresponds to evaluating all possible plans of the max fanout, i.e., using
$\lambda_{max}$ segments (cf. Figure~\ref{fig:hier}, level Hier 1 of the hierarchy), and the second term to evaluating all possible plans of smaller fanouts (cf. Figure~\ref{fig:hier}, level Hier 2).
In the leaf node packing algorithm, given that the final pack number is $np$, the time complexity of node packing is $O(2^{\lambda_N}*np)$.
In summary, the total in-core complexity is $O(\sum_N(4wc_N+\binom{w}{\lambda_{max}}*2^{w} + \sum_{i=\lambda_{min}}^{\lambda_{max}-1}\binom{w}{i}2^{\lambda_{i+1}} + 2^{\lambda_N}*np))$.

Random disk writes can have a significant cost when building Dumpy 
(cf. Figure~\ref{fig:pipeline}, Stage 5).
Assume the number of data series in the database is $|db|$, the number of leaf nodes is $n_l$ and the memory buffer can contain $B$ series.
In each batch, Dumpy generates $n_l$ random writes at most, and in total, $O(\frac{|db|}{B}*n_l)$ random writes for the whole index building.

For querying, the approximate search goes down a single path from the root node to a target leaf node. 
Let the length of this path be $|p|$; then the cost is $O(|p|w)$.
The I/O cost is a single disk read of size $O(th)$.
Compared to the iSAX indexes (with binary fanout), the length $|p|$ of the Dumpy path is $2/\overline{\lambda}$x smaller, where $\overline{\lambda}$ denotes Dumpy's average fanout.
In addition to the target leaf node search cost, the complexity of the exact search comprises of $O((1-pr)*n_l)$ random disk reads of size $O(th)$, where $pr$ is the pruning ratio, and $O(w*n_{total}\log n_{total})$ in-core calculations, where $n_{total}$ is the total number of nodes.

In practice, Dumpy is a more compact index (smaller $n_l$ and $n_{total}$ values) than other SAX-based indexes (cf. Section~\ref{sec:expr-build}), and therefore, faster in both building and querying times.

\noindent{\textbf{[Space complexity]}}
The space occupied by Dumpy (in addition to the raw data size) is as follows. 
The SAX words are persisted on disk, 
occupying $\lceil wb*|db| / 8 \rceil$ bytes.
The internal nodes of the index store the routing table, the iSAX word, and the list of segments used in the split (i.e., the chosen segments), for a total of $\sum_{N}(8*2^{\lambda_{N}} + wb/8 + \lambda_{N})$ bytes. 
The leaf nodes store a single iSAX word summary, for a total of $n_l*(wb/8)$
bytes.
Since the number of nodes is small, Dumpy introduces very little additional storage in practice.

\section{Dumpy-Fuzzy}
\label{sec:Dumpy-f}
As partition-based indexes, data series indexes also suffer from the so-called \emph{boundary issue} in approximate search~\cite{nsg,spann}.
That is, the data series located near the boundary of a query's resident node 
are also good candidates, but cannot be considered in approximate search since they may be located in different subtrees.
To overcome this problem, we propose a variant of Dumpy, named Dumpy-Fuzzy, that views the static partition boundary (i.e., the SAX breakpoints) as a range (fuzzy boundary) and places the series lying on this range into the nodes of both sides.
Dumpy-Fuzzy further improves the approximate search accuracy compared with Dumpy at the expense of a small overhead on index building and disk space. 

Specifically, Dumpy-Fuzzy adds a duplication procedure after splitting.
For each newly-generated internal node $N$, it checks the series 
lying on $N$'s neighboring nodes (i.e., the nodes whose $sid$ is $1$-bit different from $N$) and duplicates the series near the boundary into itself.
For example, a node with $sid=000$ will check the series of neighboring nodes $100$ on the first chosen segment and duplicate the series that are very close to $000$ into $000$. The same process applies to nodes $010$ and $011$ on the second and third chosen segments, respectively.

We introduce a hyper-parameter $f \in (0,1)$, the fuzzy boundary ranges regarding the original node ranges, to control which series is qualified to be duplicated.
In addition, duplication also applies after leaf node packing.
The series near the boundaries of a leaf pack can also be placed redundantly into the pack 
in the same way as above.
we ensure that no additional split will be introduced in this procedure (i.e., the leaf pack will not overflow).

Note that Dumpy-Fuzzy does not damage Dumpy's pruning power for exact search.
Duplicated series do not change the iSAX words of nodes or packs. 
Hence, the lower bound calculations are kept the same.
Therefore, without violating the pruning-based exact search, Dumpy-Fuzzy improves the approximate search accuracy
by examining more promising candidates.

We also claim that neither binary nor full-ary structures can easily adopt similar fuzzy boundary optimizations.
In a binary structure, since each leaf node has only one sibling node, it is prone to produce excessive duplication series and generate more splits.
This results in a deeper index, which rather decreases the search accuracy.
As for the full-ary structure, the excessive number of leaf nodes translates into an overwhelming number of replication destinations, introducing unacceptable storage overhead.
We have experimentally verified these observations but omit the results and the detailed algorithms due to the lack of space.

\section{Experiments}
\label{sec:expr}
\noindent\textbf{[Environment]} Experiments were conducted on an Intel Core(R) i9-10900K 2.80GHz 10-core CPU with 4*32GB 2400MHz main memory, running Windows Subsystem of Linux (Ubuntu Linux 20.04 LTS). The machine has a Samsung PCIe 2TB SSD (default), and a Seagate SATAIII 7200RPM 2TB HDD. 
Our codes are available at \url{https://github.com/DSM-fudan/Dumpy}.

\noindent\textbf{[Datasets]} 
We use one synthetic and three real datasets. All series are z-normalized before indexing and querying. 
In each dataset, we prepare 200 queries that are not part of the dataset to ensure sufficient hardness~\cite{hardness}, and obtain the ground truth kNN results using brute-force search.
\textbf{Rand} is a synthetic dataset, generated as cumulative sums of random walk steps following a standard Gaussian distribution $N(0,1)$. 
It has been extensively used in the existing works~\cite{dft, DBLP:journals/vldb/ZoumpatianosLIP18, hydra1, hydra2}. 
We generate 50-800 million Rand series of different lengths (50GB-800GB). 
\textbf{DNA}~\cite{dna} is a real dataset collected from DNA sequences of two plants, Allium sativum and Taxus wallichiana.
It comprises 26 million data series of length 1024 ($\sim$113GB).
The second real dataset, \textbf{ECG} (Electrocardiography), is extracted from the MIMIC-III Waveform Database~\cite{ecg}.
It contains over 97 million series of length 320 ($\sim$117GB), sampled at 125Hz from 6146 ICU patients.
The last real dataset, \textbf{Deep}~\cite{deep}, comprises 1 billion vectors of size 96, extracted from the last CNN (convolutional neural network) layers of images.

\noindent\textbf{[Algorithms]} 
In the iSAX-index family, we take \textbf{iSAX2+} as the SOTA binary structure~\cite{hydra2}.
We also implement a stand-alone version of \textbf{TARDIS} as the SOTA full-ary structure and use 100\% sampling percent.
\textbf{DSTree}~\cite{ds-tree} is also included as one of the SOTA data series indexes~\cite{hydra2}.
For simplicity, Dumpy-Fuzzy with parameter $f$ is abbreviated as \textbf{Dumpy-$f$}. 
To evaluate the quality of these indexes, we implement extended approximate search, as well as pruning-based search.

All the codes are open-source, implemented in C/C++, and compiled by g++ 9.4.0 with -O3 optimization. 
We use the optimized versions of DSTree and iSAX2+~\cite{hydra1}. 

\noindent\textbf{[Parameters]} We set the number of segments $w=16$, SAX cardinality $c=64$ (i.e., $b = 8$), and the leaf size threshold $th=10000$. 
The memory buffer size for index building is set to $4GB$ unless specified.
The replication times of each series in Dumpy-$f$ is set to at most 3.

\noindent\textbf{[Measures]}
Similar with other works~\cite{hydra2,hd-index,new-pq}, we use 
Mean Average Precision (MAP)~\cite{map} as the accuracy measure, which is defined as the mean value of AP on a group of queries. For query $s_q$, AP equals to
$\frac{1}{k}\sum_{i=1}^{k}P(s_q,i)*rel(i)$, where $P(s_q,i)$ is the ratio of true neighbors among the top-$i$ nearest  results and $rel(i)$ is 1 if the $i$-th nearest result is the true exact $k$NN result and 0 otherwise. It can be proved that MAP is equivalent to the average recall rate when the returned results are sorted by the actual distances.
Another similarity measure we use is the average error ratio which measures the difference between approximate and exact results, commonly used in approximate search ~\cite{hd-index, tardis}, and defined as $\frac{1}{k}\sum_{i=1}^{k}\frac{dist(a_i,s_q)}{dist(r_i,s_q)}$.
We measure both ED and DTW, where the DTW warping window size is set to $10\%$ of the series length as a common setting~\cite{messi,dtw}.

\begin{figure}[tb]
\subfigure[Building time across four datasets on SSD]{
\label{fig:cons-dataset} 
\includegraphics[width=0.5\linewidth]{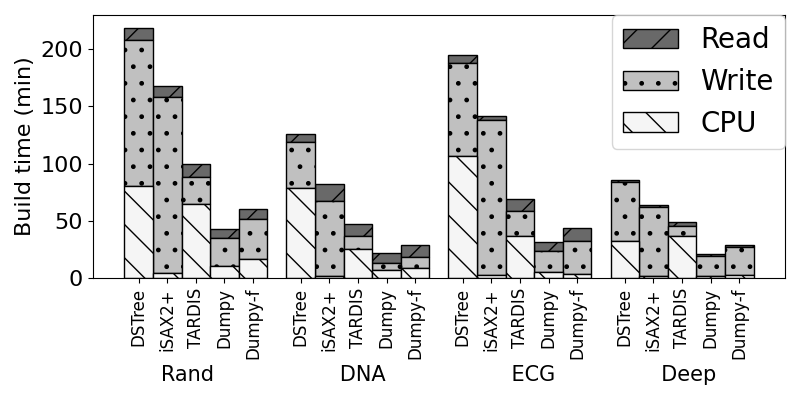}
}
\subfigure[Disk writes in building]{
\label{fig:cons-write} 
\includegraphics[width=0.24\linewidth]{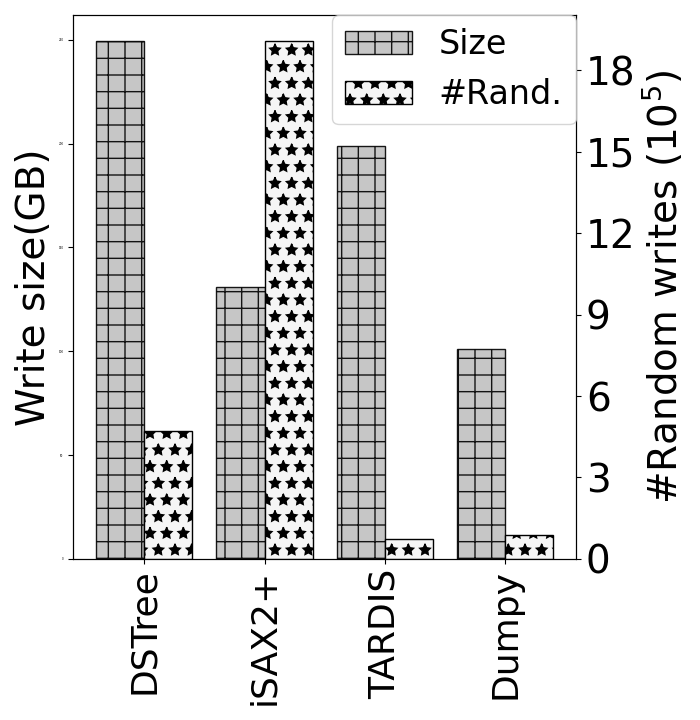}
}
\subfigure[Building time on HDD]{
\label{fig:cons-hdd} 
\includegraphics[width=0.2\linewidth]{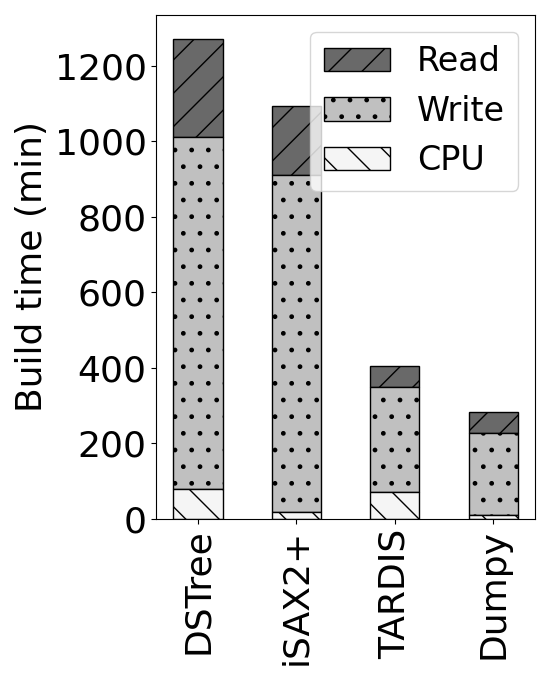}
}
\caption{Index building (4GB memory)}
\label{fig:cons} 
\end{figure}

\begin{table}[tb]
{
\scriptsize
	\centering
	\caption{Index structure statistics}
	\label{tab:structure}
	\begin{threeparttable}
	\begin{tabular}{c|c|c|c|c|c|c}
		\hline
		Dataset & Method & \#Leaves & \#Nodes  & Height &  Fill factor & {Index Size (MB)\tnote{1}}\\
		\hline\hline
		\multirow{4}{*}{Rand} & iSAX2+ & 73563 & 86945 & 20  &13.59\%  & {16}\\
		\cline{2-7}
		&                 DSTree & 17847 & 35693 & 32  & 56.03\%  & {9}\\
		\cline{2-7}
		&                TARDIS   &8516867  & 8520065 & \textbf{3} & 0.11\% & {732}\\
		\cline{2-7}
		&                \textbf{Dumpy}   & \textbf{14106} &  \textbf{19418} & 7 & \textbf{70.89\%}  & {\textbf{3}}\\
		\hline\hline
		\multirow{4}{*}{DNA}  & iSAX2+ & 42906 & 47885 & 25  & 6.14\%  & {9}\\
		\cline{2-7}
		&         DSTree &  5833& 11665 &  43 & 45.16\%  & {3}\\
		\cline{2-7}
		&         TARDIS   &  1011436 & 1312989  & \textbf{5} & 0.26\% & {278}\\
		\cline{2-7}
		&         \textbf{Dumpy}   & \textbf{4367} &  \textbf{6228} & 9 & \textbf{60.32\%} & {\textbf{1}}\\
		\hline\hline
		\multirow{4}{*}{ECG}  & iSAX2+ & 69786 & 74042 & 9  & 13.98\%  & {14}\\
		\cline{2-7}
		&     DSTree & 20740 & 41479 & 48  & 47.04\% & {13}\\
		\cline{2-7}
		&     TARDIS   & 3178628 &3182368  &\textbf{4}  & 0.33\% & {749}\\
		\cline{2-7}
		&     \textbf{Dumpy}   & \textbf{12112} & \textbf{15050} & 7 & \textbf{80.55\%} & {\textbf{3}} \\
		
		\hline\hline
		\multirow{4}{*}{Deep}  & iSAX2+ & 68096 & 71188 & 8  & 19.08\% & {11}\\
		\cline{2-7}
		&     DSTree & 16324 & 32647 & 33  & 61.26\% & {8}\\
		\cline{2-7}
		&     TARDIS   & 824458  & 827094 & \textbf{3} & 0.27\% & {546}\\
		\cline{2-7}
		&     \textbf{Dumpy}   & \textbf{11590} & \textbf{13664} & 8 & \textbf{86.28\%} & { \textbf{3}} \\
		
		\hline\hline
		
	\end{tabular}
	\begin{tablenotes} 
        \footnotesize   
        {
        \item[1] Size of in-memory index structure only.  
        }
      \end{tablenotes}         
    \end{threeparttable}
    } 
\end{table}

\subsection{Index Building}
\label{sec:expr-build}
\subsubsection{Efficiency}
First, we evaluate the index building efficiency in four datasets on SSD, and the results are shown in Figure~\ref{fig:cons-dataset}. 
In all four datasets, Dumpy outperforms the other three methods by a large margin, i.e, $5.3$ times faster than DSTree, $3.8$ times than iSAX2+ and $2.5$ times than TARDIS on average. 
Dumpy-$f$ only incurs small overheads (about \textbf{38\%}) on Dumpy and is considerably faster than DSTree and iSAX2+.
The advantage of Dumpy mainly comes from the reduction of random  disk writes. 
As shown in Figure~\ref{fig:cons-write}, the number of random writes (\#Rand.) of iSAX2+ is $\sim$\textbf{3.5}x more than DSTree and $\sim$\textbf{17}x more than Dumpy and TARDIS, making it the dominating factor in index building time than the number of writing bytes (\#Seq.) in both SSD and HDD.
Though TARDIS also enjoys low I/O costs for its compact and large partition, it needs more CPU time to serialize the enormous index structure into each partition. Results on HDD in Figure~\ref{fig:cons-hdd} are similar to those on SSD.

We present the detailed index information in Table~\ref{tab:structure}. Dumpy has the fewest leaf nodes (i.e., the highest fill factor) which verifies the good compactness of Dumpy. That of DSTree is slightly larger than Dumpy, and that of iSAX2+ is generally $>$\textbf{3}x more than DSTree.
It verifies the complexity analysis in Section~\ref{sec:workflow}.
Noting that the index building time of DSTree is even longer than iSAX2+ (Figure~\ref{fig:cons-dataset}), due to its high CPU cost and large writing bytes incurred by EAPCA calculations.
TARDIS generates million-level leaves and has a low fill factor for its full-ary structure. 
These nodes are organized in large partitions where each partition is 128MB, as the setting of ~\cite{tardis}.

\begin{figure}[tb]
\includegraphics[width=0.6\linewidth]{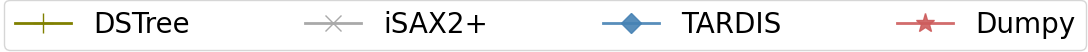} \\
\subfigure[On data series length]{
  \includegraphics[width=0.35\linewidth]{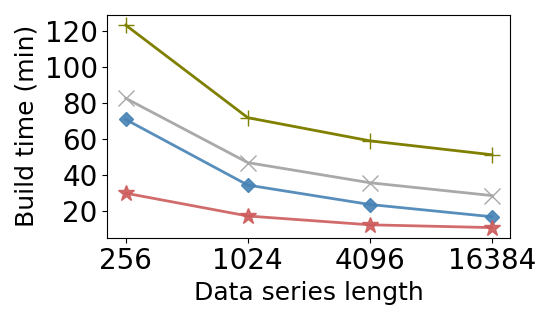}
  \label{fig:cons-length}
}
\subfigure[On dataset size]{
\includegraphics[width=0.37\linewidth]{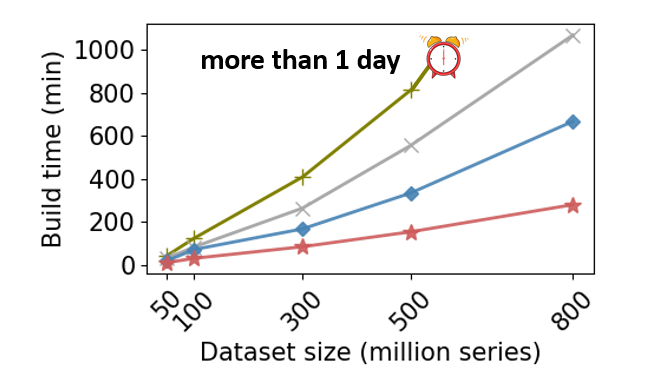}
\label{fig:cons-size}
}
\caption{Index scalability (32GB memory)}
\label{fig:scale} 
\end{figure}

\begin{figure*}[tb]
\includegraphics[width=0.96\linewidth]{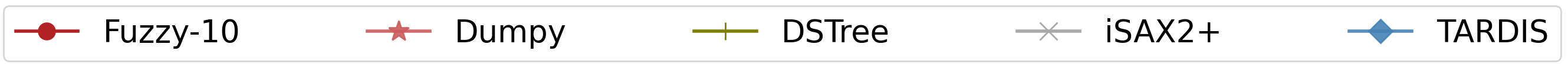} 

\begin{minipage}{0.96\linewidth}
    \subfigure[Rand]{
	\begin{minipage}[t]{0.3\linewidth}
		\centering
		\includegraphics[width=\linewidth]{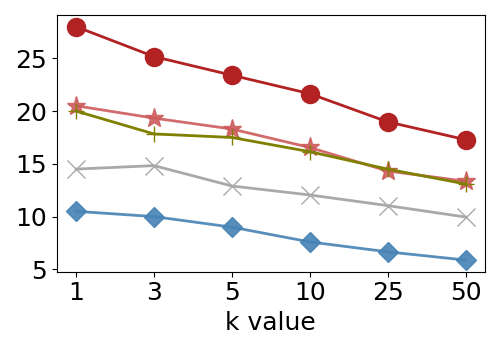}
		\label{fig:rand-recall}
	\end{minipage}%
	}
	\subfigure[DNA]{
	\begin{minipage}[t]{0.3\linewidth}
		\centering	
		\includegraphics[width=\linewidth]{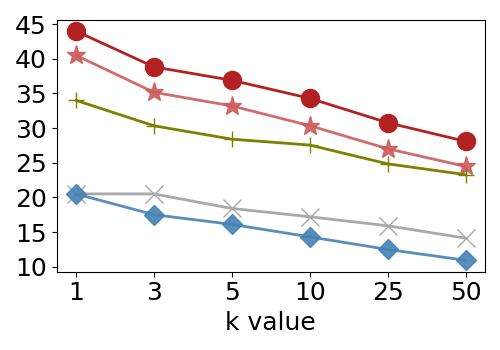}
		\label{fig:dna-recall}
	\end{minipage}%
	}
	\subfigure[ECG]{
	\begin{minipage}[t]{0.3\linewidth}
		\centering
		\includegraphics[width=\linewidth]{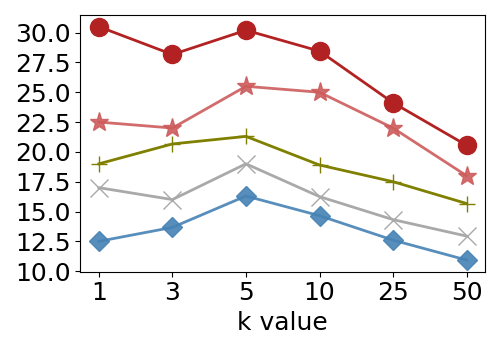}
		\label{fig:ecg-recall}
	\end{minipage}
	}

 \subfigure[Rand]{
	\begin{minipage}[t]{0.3\linewidth}
		\centering
		\includegraphics[width=\linewidth]{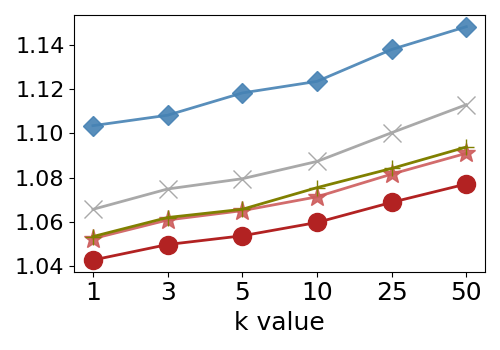}
		\label{fig:rand-precision}
	\end{minipage}%
	}
	\subfigure[DNA]{
	\begin{minipage}[t]{0.3\linewidth}
		\centering	
		\includegraphics[width=\linewidth]{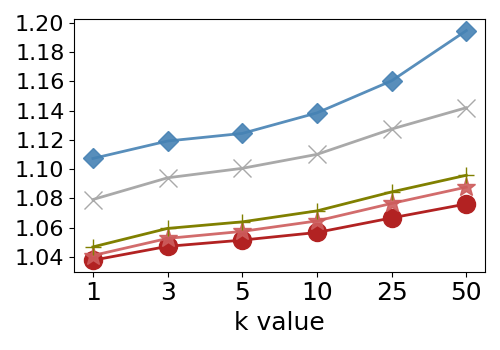}
		\label{fig:dna-precision}
	\end{minipage}%
	}
	\subfigure[ECG]{
	\begin{minipage}[t]{0.3\linewidth}
		\centering
		\includegraphics[width=\linewidth]{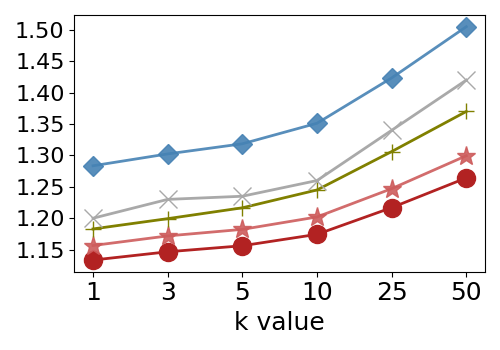}
		\label{fig:ecg-precision}
	\end{minipage}
	}
\caption{Approximate search under ED (search one node).}
\label{fig:approx} 
\end{minipage}
\begin{minipage}{0.96\linewidth}
    \subfigure[Rand]{
	\begin{minipage}[t]{0.3\linewidth}
		\centering
		\includegraphics[width=\linewidth]{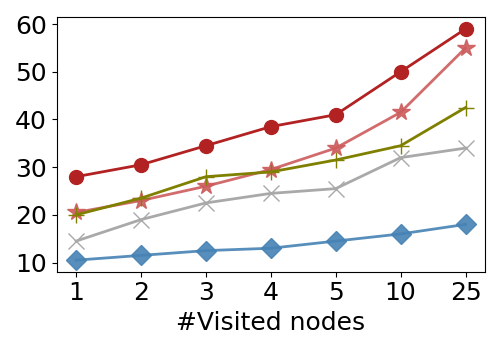}
		\label{fig:rand-node-recall}
	\end{minipage}%
	}
	\subfigure[DNA]{
	\begin{minipage}[t]{0.3\linewidth}
		\centering	
		\includegraphics[width=\linewidth]{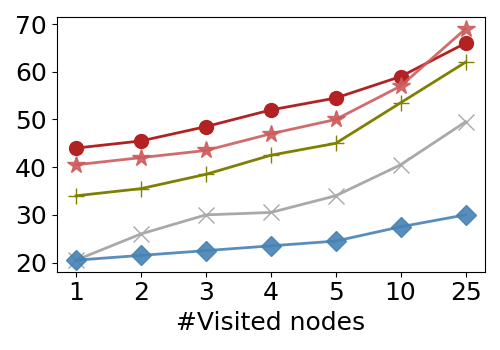}
		\label{fig:dna-node-recall}
	\end{minipage}%
	}
	\subfigure[ECG]{
	\begin{minipage}[t]{0.3\linewidth}
		\centering
		\includegraphics[width=\linewidth]{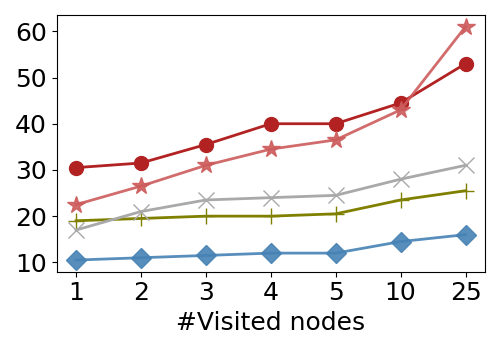}
		\label{fig:ecg-node-recall}
	\end{minipage}
	}
	
	\subfigure[Rand]{
	\begin{minipage}[t]{0.3\linewidth}
		\centering
		\includegraphics[width=\linewidth]{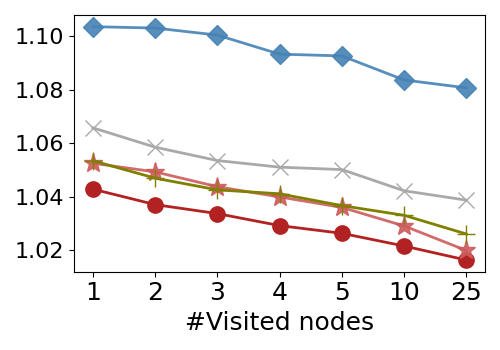}
		\label{fig:rand-node-precision}
	\end{minipage}%
	}
	\subfigure[DNA]{
	\begin{minipage}[t]{0.3\linewidth}
		\centering	
		\includegraphics[width=\linewidth]{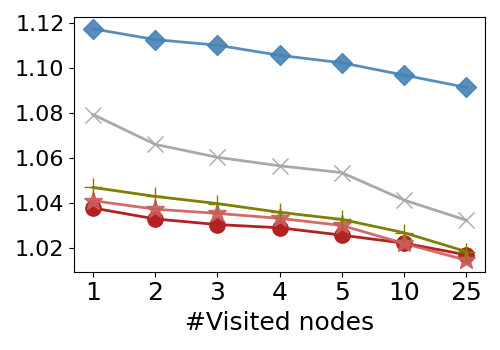}
		\label{fig:dna-node-precision}
	\end{minipage}%
	}
	\subfigure[ECG]{
	\begin{minipage}[t]{0.3\linewidth}
		\centering
		\includegraphics[width=\linewidth]{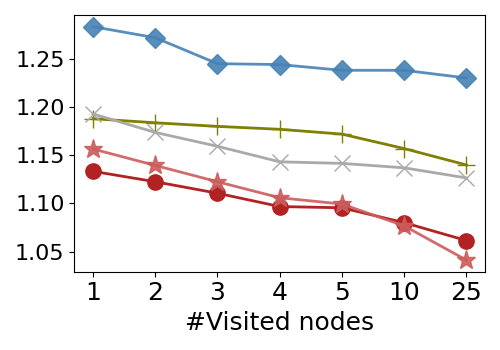}
		\label{fig:ecg-node-precision}
	\end{minipage}
	}
\caption{Extended approximate search under ED (k=1).}
\label{fig:approx-node} 
\end{minipage}
\end{figure*}

\begin{figure}[tb]
\begin{minipage}{.5\linewidth}
    \includegraphics[width=\linewidth]{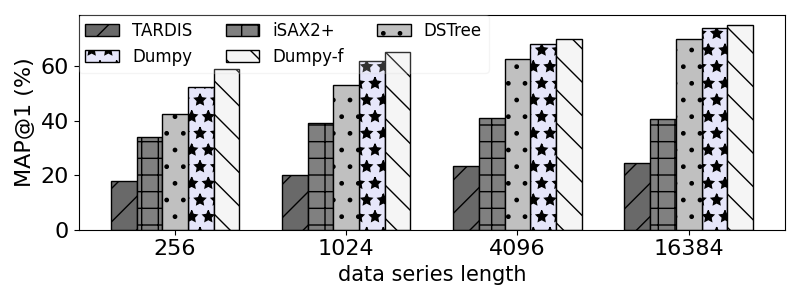}
    \captionof{figure}{Extended approx. Search vs. series lengths (search 25 nodes)}
    \label{fig:app-length}
\end{minipage}%
\begin{minipage}{.24\linewidth}
    \includegraphics[width=\linewidth]{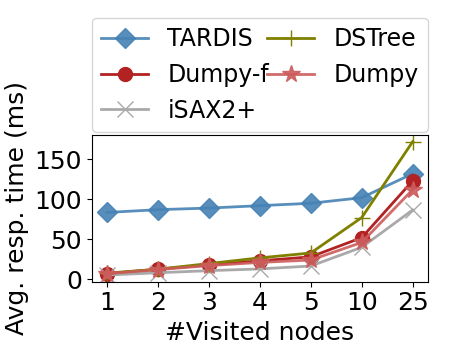}
    \captionof{figure}{Query time}
    \label{fig:qps}
\end{minipage}%
\begin{minipage}{.24\linewidth}
    \centering
    \includegraphics[width=\linewidth]{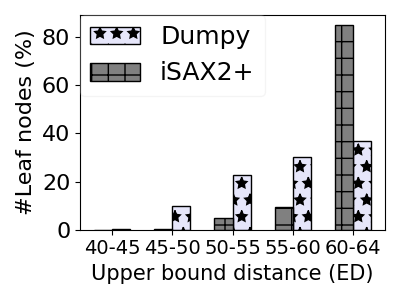}
    \captionof{figure}{ Histogram of upper bounds of the distance}    
    \label{fig:accuracy-bound}
\end{minipage}%
\end{figure}

\begin{figure}[tb]
\includegraphics[width=0.96\linewidth]{legend.png} 

\begin{minipage}{.48\linewidth}
    \subfigure[$k=1$]{
	\begin{minipage}[t]{0.48\linewidth}
		\centering
		\includegraphics[width=\linewidth]{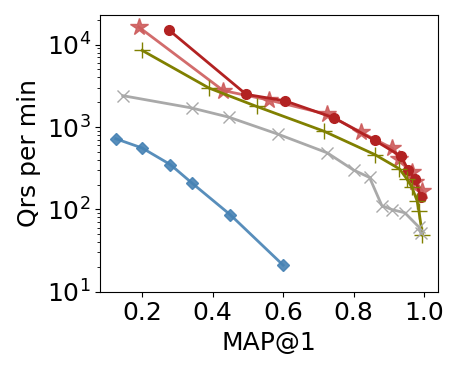}
	\end{minipage}%
	}
    \subfigure[$k=50$]{
	\begin{minipage}[t]{0.48\linewidth}
		\centering
		\includegraphics[width=\linewidth]{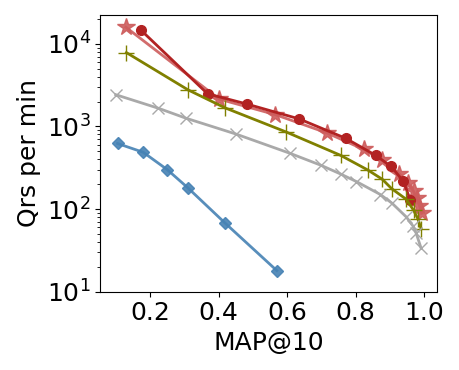}
	\end{minipage}%
	}
    \captionof{figure}{Efficiency v.s. accuracy}
    \label{fig:ng}
\end{minipage}%
\begin{minipage}{.48\linewidth}
    \subfigure[MAP]{
	\begin{minipage}[t]{0.48\linewidth}
		\centering
		\includegraphics[width=\linewidth]{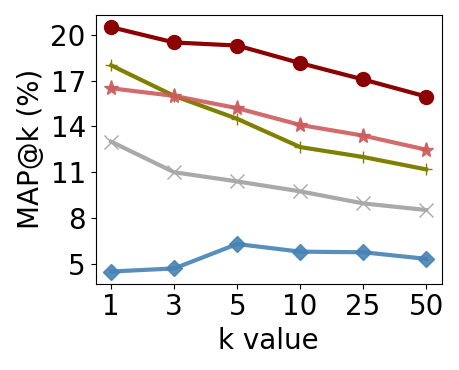}
	\end{minipage}%
	}
    \subfigure[Avg. error ratio]{
	\begin{minipage}[t]{0.48\linewidth}
		\centering
		\includegraphics[width=\linewidth]{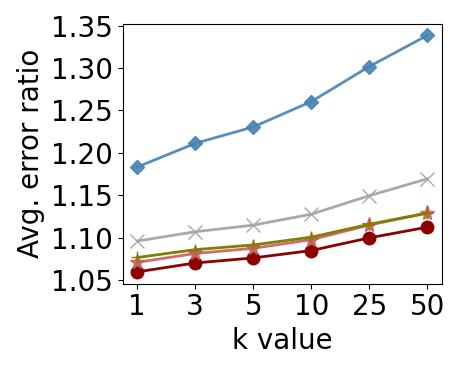}
	\end{minipage}%
	}
    \captionof{figure}{Approx. Search under DTW}
    \label{fig:dtw}
\end{minipage}%
\end{figure}

\begin{figure}[tb]
\begin{minipage}{.48\linewidth}
    \subfigure[Effect of $w$]{
	\begin{minipage}[t]{0.6\linewidth}
		\centering
		\includegraphics[width=\linewidth]{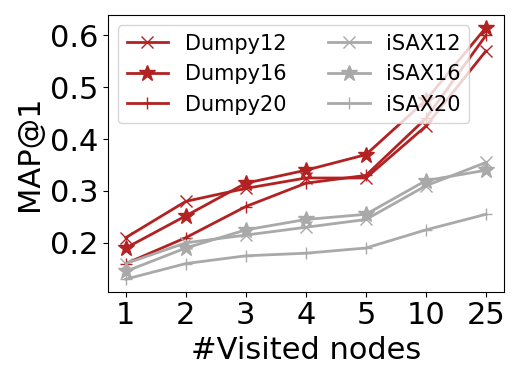}
        \label{fig:seg}
	\end{minipage}%
	}
    \subfigure[Effect of $\alpha$]{
	\begin{minipage}[t]{0.35\linewidth}
		\centering
		\includegraphics[width=\linewidth]{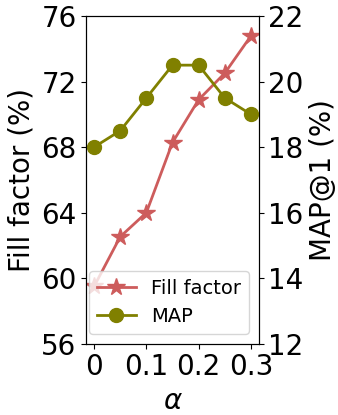}
        \label{fig:alpha}
	\end{minipage}%
	}
    \captionof{figure}{Influence of parameters}
    \label{fig:parameter}
\end{minipage}%
\begin{minipage}{.48\linewidth}
    \subfigure[On index building]{
	\begin{minipage}[t]{0.48\linewidth}
		\centering
		\includegraphics[width=\linewidth]{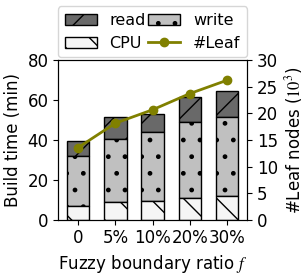}
        \label{fig:fuzzy-1}
	\end{minipage}%
	}
    \subfigure[On Search Accuracy]{
	\begin{minipage}[t]{0.48\linewidth}
		\centering
		\includegraphics[width=\linewidth]{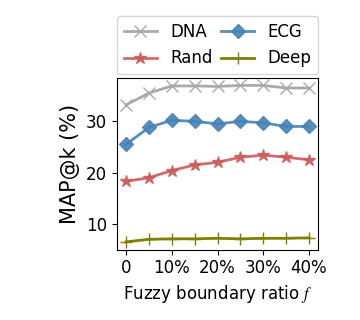}
        \label{fig:fuzzy-2}
	\end{minipage}%
	}
    \captionof{figure}{Influence of $f$}
    \label{fig:fuzzy}
\end{minipage}%
\end{figure}

\subsubsection{Scalability}
Next, we test the scalability in Rand datasets by increasing data size from 50GB to 800GB, and series length from 256 to 16384.
When the dataset size is varied, the series length is kept constant at 256, whereas the dataset size is kept at 100GB when the length is varied, as the same design with the benchmark~\cite{hydra1}.

Figure~\ref{fig:scale} presents the index building time with 32GB memory. 
Dumpy has the best scalability in both cases.
In a linear regression test for the building time and dataset size, Dumpy's coefficient of determination $R^2$ is \textbf{0.9904}, verifying its linear growth of the building time.
The reason is that the number of leaves increases linearly as the dataset scales up, indicating a nearly constant average fill factor.
This also supports the complexity analysis.
The performance when varying series length also follows this rule.

\subsection{Query Processing}
In this section, we verify the accuracy of the query processing.

\subsubsection{Approximate Search}
\label{sec:expr-approx-search}
First, we evaluate the accuracy of approximate results across four datasets. 

\noindent\textbf{[Search one node]} 
We compare these approaches when searching one node to obtain the approximate top-$k$ result on three datasets with ED distance, and results are shown in Figure~\ref{fig:approx}. It can be seen that Dumpy consistently outperforms other approaches.
Specifically, Dumpy improves the average accuracy by \textbf{84}\%, \textbf{46}\%, \textbf{11}\% and reduces the average error ratio by \textbf{7.3}\%, \textbf{3.4}\% and  \textbf{1.4}\% on three datasets compared with TARDIS, iSAX2+ and DSTree respectively. TARDIS has the lowest performance, due to its low fill factor. 
iSAX2+ suffers from insufficient intra-node series proximity, characterized by the number of uneven nodes as described in Figure~\ref{fig:heuristic} ($>$\textbf{20}\% leaf nodes have one segment using more than 4 bits than other segments).
Moreover, Dumpy-Fuzzy has higher accuracy than Dumpy and other approaches, which verifies our duplication strategy.

\noindent\textbf{[Search multiple nodes]} 
In Figure~\ref{fig:approx-node}, we compare the accuracy of searching multiple nodes (1 to 25) for top-1 result with ED distance.
The MAP value of Dumpy and Dumpy-$f$ increases remarkably faster than the competitors, attributed to our multi-ary structure that provides closer sibling nodes.
When visiting 25 nodes, Dumpy and Dumpy-$f$ improve the accuracy by \textbf{58}\%, \textbf{65}\% and reduce the average error ratio by \textbf{3.6}\% and \textbf{3.7}\% on average of four datasets respectively compared with the second-best approach, DSTree.
We also compare the accuracy as the series length varies (Figure~\ref{fig:app-length}). The accuracy on different lengths shares similar rankings.

Figure~\ref{fig:qps} shows the average query time of different approaches for approximate top-50 queries. It can be seen that Dumpy has a similar query processing time with iSAX2+ and is faster than DSTree. That is, our approach can achieve higher accuracy with a smaller time cost.
TARDIS is much slower because it pays lots of time to deserialize the large index partition. 

Our results show that Dumpy can achieve 60\%-70\% MAP within 100ms on TB-level datasets, while its index building time is 4x faster than the SOTA competitors. 
These results demonstrate that Dumpy fulfills the requirements of many kNN-based applications.

\noindent\textbf{[Upper bound of approximate NN distance]} 
Dumpy (like any other index) does not provide any accuracy guarantees for its (ng-)approximate search results~\cite{hydra1,hydra2}. Nevertheless, we can provide an upper bound for the distance between the approximate NN and the query series. 
This upper bound can be interpreted as the worst-case accuracy of the approximate search results of Dumpy. 

Given the target leaf node, this upper bound distance is defined as $\sqrt{\frac{l}{w}\sum_{i=1}^w range_i^2}$, where $range_i$ is the distance between two breakpoints in the $i$-th segment. It corresponds to the distance when the query and the nearest neighbor series are located \emph{on} the opposite boundaries of each segment.

In Figure~\ref{fig:accuracy-bound}, we use a histogram to show the distribution of these upper bound distances for the leaf nodes of the Dumpy and iSAX indexes.
The histogram shows that over \textbf{80}\% of the iSAX leaf nodes have loose bounds (60-64), whereas over \textbf{60\%} of the Dumpy leaves have tight bounds (40-60).
This is because Dumpy chooses to split the coarser segment, which leads to a smaller worst-case distance than iSAX.

\noindent\textbf{[Efficiency vs accuracy]} 
We extend the approximate search to all leaf nodes with lower bound pruning to evaluate the indexes' response time under the whole MAP range.
The results are depicted in Figure~\ref{fig:ng}.
Benefiting from the high-proximity nodes and compact index structure,
Dumpy surpasses its competitors in both low and high-precision intervals.

\noindent\textbf{[Searching under DTW]} 
In this experiment, we compare the accuracy under DTW distance in Figure~\ref{fig:dtw}.
Due to the inherent hardness, the precision is lower than ED generally.
However, Dumpy and Dumpy-$f$ still achieve better precision and error ratio under DTW.
Since the absolute distance of DTW is smaller than ED, the differences in the error ratio among all the methods tend to be smaller (except for TARDIS).

\noindent\textbf{[Influence of parameters]} 
We also investigate the influence of parameters on the accuracy, including {the number of segments $w$,} the weight factor $\alpha$ in the objective function Equation.~\ref{equ:score} and boundary ratio $f$ for Dumpy-Fuzzy.

{In Figure~\ref{fig:seg}, we vary $w$ for iSAX and Dumpy and observe that the best iSAX performance (when $w=16$)
is still worse than 
Dumpy (with any number of segments $w \in \{12, 16, 20\}$).
This is explained by Dumpy's strategy to search for better-quality splits that fully exploit the iSAX summarization power of the available segments.
We also observe that in both indexes, fewer segments tend to degrade the precision of iSAX summarizations, whereas more segments produce 
a large number of leaf
nodes, which harms the index performance.
}

In Figure~\ref{fig:alpha}, 
when $\alpha$ increases from 0 to 0.3, the fill factor also keeps increasing, from 59\% to 75\%.
The precision first increases since visiting more series in a node, and then decreases due to the reduction of intra-node proximity.
As shown in Figure~\ref{fig:heuristic}, when $\alpha$ decreases, the split plan tends to be like Figure~\ref{fig:heuristic}(c), which means that the series in a node is not that similar.
The sweet point for the precision occurs when $\alpha$ is about 0.2, which is used across different datasets in our experiments, showing the robust performance. 

Second, we evaluate the stability of Dumpy-Fuzzy regarding different $f$, the fuzzy boundary ratio.
In Figure~\ref{fig:fuzzy-1} we observe that the number of leaves and the building time increases slowly as $f$ increases. 
In Figure~\ref{fig:fuzzy-2} ($k=5$), the precision and error ratio improve with $f$ increasing to \textbf{10}\%, and remain relatively stable then.
It implies that Dumpy-Fuzzy is stable regarding $f$, and choose $f$=10 for ECG and Deep and $f$=30 for Rand and DNA.

\subsubsection{Exact Search}

We evaluate the exact search efficiency of Dumpy against other methods in Table~\ref{tab:exact}.
Since TARDIS does not support exact $k$NN search in the original paper, we implement a similar algorithm as the iSAX-index family.
But only the node summarized with iSAX words can be pruned during searching.
The results are reported using the average of 40 queries with $k=50$ under ED and DTW.
Overall, Dumpy achieves the best efficiency in all cases. It is worth noting that although DSTree has a higher pruning ratio than Dumpy, the response time is still slower than Dumpy. The reason is as follows. DSTree takes a longer time to compute the lower bound of distance due to computing the standard deviation frequently. iSAX2+ suffers from the low fill factor and needs to read about \textbf{3} times nodes more than Dumpy and DSTree.

\begin{table}[bt]
{
\scriptsize
	\centering
	\caption{Exact Search. 
 For response time,
 the first number in the brackets is the I/O time while the second is the CPU time.}
	\label{tab:exact}
	\begin{tabular}{c|c|c|c|c}
		\hline
		    & Method & Resp. time (s) & \#Loaded Nodes   & Pruning ratio \\
		\hline\hline
		\multirow{4}{*}{Rand-ED} & iSAX2+ & 65 (50+15) & 7595   &81.51\% \\
		\cline{2-5}
		&                 DSTree & 33 (20+13) & 2027   & 86.06\%\\
		\cline{2-5}
		&                TARDIS   & 53 (15+38) & 1665744 & 59.30\%\\
		\cline{2-5}
		& \textbf{Dumpy}   & \textbf{17} (\textbf{12}+\textbf{5}) & 2641 & 83.70\%\\
		\hline\hline
		\multirow{4}{*}{Rand-DTW}  &iSAX2+ & 151 (85+66) & 18660   & 72.58\% \\
		\cline{2-5}
		&         DSTree & 79 (24+65) & 3678  & 75.12\%\\
		\cline{2-5}
		&         TARDIS   & 208 (22+186) & 2846868  &  49.24\% \\
		\cline{2-5}
		&         \textbf{Dumpy}   & \textbf{58} (\textbf{18}+\textbf{40}) & 3997  & 73.61\% \\
		\hline\hline
		\multirow{4}{*}{DNA-ED} & iSAX2+ & 42 (26+16) & 1077   &91.04\% \\
		\cline{2-5}
		&     DSTree & 21 (16+5) & 326   & 94.00\%\\
		\cline{2-5}
		&          TARDIS   & 40 (11+29) & 161959 & 71.64\%\\
		\cline{2-5}
		& \textbf{Dumpy}   & \textbf{12} (\textbf{10}+\textbf{2}) & 433 & 91.69\%\\
		\hline\hline
		\multirow{4}{*}{DNA-DTW}  & iSAX2+ & 116 (60+56)& 2163 & 87.77\% \\
		\cline{2-5}
		&         DSTree & 63 (18+45) & 497  & 90.93\%\\
		\cline{2-5}
		&         TARDIS   & 143 (16+127) & 194645 & 68.78\% \\
		\cline{2-5}
		&         \textbf{Dumpy}   & \textbf{53} (\textbf{14}+\textbf{39}) & 528 & 89.41\% \\
		\hline\hline
	\end{tabular}
    } 
\end{table}

\subsection{Complete Workloads}
\begin{figure}[tb]
  \includegraphics[width=0.7\linewidth]{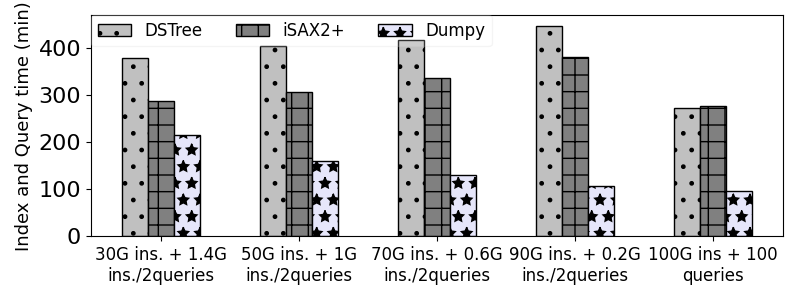}
  \caption{Update performance (4GB memory)}
  \label{fig:conmplte-workload}
\end{figure}
Finally, we compare different approaches when inserting new data series (Figure~\ref{fig:conmplte-workload}).
We omit TARDIS since it is designed for the static dataset and not easy to be extended.
To be fair, we implement all methods using a single thread (even though Dumpy is multi-threaded).
We use different synthetic workloads consisting of 100 exact queries, and a total of 100 million series, where queries are interleaved by a batch of insertions.
The results show Dumpy outperforms the competitors for all workloads, thanks to its compact structure.
Although the re-splitting and re-packing procedures add an additional cost, this cost is balanced by the efficiency improvements that these two designs bring along.
Moreover, Dumpy shows better performance when the initial batch size increases (while iSAX and DSTree show worse performance), because fewer insertions incur fewer re-splitting and re-packing actions.

\section{Conclusions and Future Work}
\label{sec:conclusion}
We propose a novel multi-ary data series index Dumpy with an adaptive split strategy that hits the right balance in the proximity-compactness trade-off.
A variant of Dumpy, Dumpy-Fuzzy can achieve even higher accuracy by clever partial duplication.
Experiments with a variety of large, synthetic and real datasets demonstrate the efficiency, scalability, and accuracy of our solutions.

In future work, we plan to extend Dumpy  
to support 
subsequence matching.
By designing a proper cost function and an efficient evaluation algorithm, 
Dumpy's adaptive splitting strategy can enhance the SOTA subsequence matching index, ULISSE~\cite{DBLP:journals/vldb/LinardiP20},  with higher-proximity nodes and hence better performance.
Moreover, by absorbing the parallel paradigms of ParIS~\cite{paris+}, SING~\cite{sing} and TARDIS~\cite{tardis}, Dumpy's overall performance can be further improved by exploiting modern hardware.

\begin{acks}
This work is supported by the Ministry of Science and Technology of China, National Key Research and Development Program (No. 2020YFB1710001).
Qitong Wang is supported by China Scholarship Council.
We would also like to thank Prof. Karima Echihabi for her kind sharing of the optimized implementation of iSAX2+ and DSTree.
\end{acks}

\bibliographystyle{ACM-Reference-Format}
\bibliography{pacmmod}


\begin{thebibliography}{74}


\ifx \showCODEN    \undefined \def \showCODEN     #1{\unskip}     \fi
\ifx \showDOI      \undefined \def \showDOI       #1{#1}\fi
\ifx \showISBNx    \undefined \def \showISBNx     #1{\unskip}     \fi
\ifx \showISBNxiii \undefined \def \showISBNxiii  #1{\unskip}     \fi
\ifx \showISSN     \undefined \def \showISSN      #1{\unskip}     \fi
\ifx \showLCCN     \undefined \def \showLCCN      #1{\unskip}     \fi
\ifx \shownote     \undefined \def \shownote      #1{#1}          \fi
\ifx \showarticletitle \undefined \def \showarticletitle #1{#1}   \fi
\ifx \showURL      \undefined \def \showURL       {\relax}        \fi
\providecommand\bibfield[2]{#2}
\providecommand\bibinfo[2]{#2}
\providecommand\natexlab[1]{#1}
\providecommand\showeprint[2][]{arXiv:#2}

\bibitem[dna({[n.\,d.]})]%
        {dna}
 \bibinfo{year}{[n.\,d.]}\natexlab{}.
\newblock \bibinfo{title}{National Center for Biotechnology Information
  (NCBI)[Internet]}.
\newblock \bibinfo{howpublished}{\url{https://www.ncbi.nlm.nih.gov/} Accessed
  March 14, 2022}.
\newblock


\bibitem[Agrawal et~al\mbox{.}(1993)]%
        {dft}
\bibfield{author}{\bibinfo{person}{Rakesh Agrawal}, \bibinfo{person}{Christos
  Faloutsos}, {and} \bibinfo{person}{Arun Swami}.}
  \bibinfo{year}{1993}\natexlab{}.
\newblock \showarticletitle{Efficient similarity search in sequence databases}.
  In \bibinfo{booktitle}{\emph{International conference on foundations of data
  organization and algorithms}}. Springer, \bibinfo{pages}{69--84}.
\newblock


\bibitem[Anagnostou et~al\mbox{.}(2020)]%
        {knn-classifier}
\bibfield{author}{\bibinfo{person}{Panagiotis Anagnostou},
  \bibinfo{person}{Petros Barbas}, \bibinfo{person}{Aristidis~G. Vrahatis},
  {and} \bibinfo{person}{Sotiris~K. Tasoulis}.}
  \bibinfo{year}{2020}\natexlab{}.
\newblock \showarticletitle{Approximate kNN Classification for Biomedical
  Data}. In \bibinfo{booktitle}{\emph{2020 IEEE International Conference on Big
  Data (Big Data)}}. \bibinfo{pages}{3602--3607}.
\newblock
\urldef\tempurl%
\url{https://doi.org/10.1109/BigData50022.2020.9378126}
\showDOI{\tempurl}


\bibitem[Arora et~al\mbox{.}(2018)]%
        {hd-index}
\bibfield{author}{\bibinfo{person}{Akhil Arora}, \bibinfo{person}{Sakshi
  Sinha}, \bibinfo{person}{Piyush Kumar}, {and} \bibinfo{person}{Arnab
  Bhattacharya}.} \bibinfo{year}{2018}\natexlab{}.
\newblock \showarticletitle{HD-Index: Pushing the Scalability-Accuracy Boundary
  for Approximate kNN Search in High-Dimensional Spaces}.
\newblock \bibinfo{journal}{\emph{Proceedings of the VLDB Endowment}}
  \bibinfo{volume}{11}, \bibinfo{number}{8} (\bibinfo{year}{2018}).
\newblock


\bibitem[Azizi et~al\mbox{.}(2023)]%
        {elpis}
\bibfield{author}{\bibinfo{person}{Ilias Azizi}, \bibinfo{person}{Karima
  Echihabi}, {and} \bibinfo{person}{Themis Palpanas}.}
  \bibinfo{year}{2023}\natexlab{}.
\newblock \showarticletitle{{ELPIS: Graph-Based Similarity Search for Scalable
  Data Science}}.
\newblock \bibinfo{journal}{\emph{Proc. {VLDB} Endow.}} \bibinfo{volume}{16},
  \bibinfo{number}{6} (\bibinfo{year}{2023}).
\newblock


\bibitem[Babenko and Lempitsky(2014)]%
        {inverted}
\bibfield{author}{\bibinfo{person}{Artem Babenko} {and} \bibinfo{person}{Victor
  Lempitsky}.} \bibinfo{year}{2014}\natexlab{}.
\newblock \showarticletitle{The inverted multi-index}.
\newblock \bibinfo{journal}{\emph{IEEE transactions on pattern analysis and
  machine intelligence}} \bibinfo{volume}{37}, \bibinfo{number}{6}
  (\bibinfo{year}{2014}), \bibinfo{pages}{1247--1260}.
\newblock


\bibitem[Babenko and Lempitsky(2016)]%
        {approx1}
\bibfield{author}{\bibinfo{person}{Artem Babenko} {and} \bibinfo{person}{Victor
  Lempitsky}.} \bibinfo{year}{2016}\natexlab{}.
\newblock \showarticletitle{Efficient indexing of billion-scale datasets of
  deep descriptors}. In \bibinfo{booktitle}{\emph{Proceedings of the IEEE
  Conference on Computer Vision and Pattern Recognition}}.
  \bibinfo{pages}{2055--2063}.
\newblock


\bibitem[Bagnall et~al\mbox{.}(2019)]%
        {DBLP:journals/dagstuhl-reports/BagnallCPZ19}
\bibfield{author}{\bibinfo{person}{Anthony~J. Bagnall},
  \bibinfo{person}{Richard~L. Cole}, \bibinfo{person}{Themis Palpanas}, {and}
  \bibinfo{person}{Konstantinos Zoumpatianos}.}
  \bibinfo{year}{2019}\natexlab{}.
\newblock \showarticletitle{Data Series Management (Dagstuhl Seminar 19282)}.
\newblock \bibinfo{journal}{\emph{Dagstuhl Reports}} \bibinfo{volume}{9},
  \bibinfo{number}{7} (\bibinfo{year}{2019}), \bibinfo{pages}{24--39}.
\newblock
\urldef\tempurl%
\url{https://doi.org/10.4230/DagRep.9.7.24}
\showDOI{\tempurl}


\bibitem[Beis and Lowe(1997)]%
        {kd-tree}
\bibfield{author}{\bibinfo{person}{J.S. Beis} {and} \bibinfo{person}{D.G.
  Lowe}.} \bibinfo{year}{1997}\natexlab{}.
\newblock \showarticletitle{Shape indexing using approximate nearest-neighbour
  search in high-dimensional spaces}. In \bibinfo{booktitle}{\emph{Proceedings
  of IEEE Computer Society Conference on Computer Vision and Pattern
  Recognition}}. \bibinfo{publisher}{IEEE}, \bibinfo{address}{San Juan, PR,
  USA}, \bibinfo{pages}{1000--1006}.
\newblock


\bibitem[Boniol et~al\mbox{.}(2020)]%
        {app2}
\bibfield{author}{\bibinfo{person}{Paul Boniol}, \bibinfo{person}{Michele
  Linardi}, \bibinfo{person}{Federico Roncallo}, {and} \bibinfo{person}{Themis
  Palpanas}.} \bibinfo{year}{2020}\natexlab{}.
\newblock \showarticletitle{Automated anomaly detection in large sequences}. In
  \bibinfo{booktitle}{\emph{2020 IEEE 36th International Conference on Data
  Engineering (ICDE)}}. IEEE, \bibinfo{pages}{1834--1837}.
\newblock


\bibitem[Boniol and Palpanas(2020)]%
        {app3}
\bibfield{author}{\bibinfo{person}{Paul Boniol} {and} \bibinfo{person}{Themis
  Palpanas}.} \bibinfo{year}{2020}\natexlab{}.
\newblock \showarticletitle{Series2graph: Graph-based subsequence anomaly
  detection for time series}.
\newblock \bibinfo{journal}{\emph{Proceedings of the VLDB Endowment}}
  \bibinfo{volume}{13}, \bibinfo{number}{12} (\bibinfo{year}{2020}),
  \bibinfo{pages}{1821--1834}.
\newblock


\bibitem[Camerra et~al\mbox{.}(2010)]%
        {isax2.0}
\bibfield{author}{\bibinfo{person}{Alessandro Camerra}, \bibinfo{person}{Themis
  Palpanas}, \bibinfo{person}{Jin Shieh}, {and} \bibinfo{person}{Eamonn
  Keogh}.} \bibinfo{year}{2010}\natexlab{}.
\newblock \showarticletitle{iSAX 2.0: Indexing and mining one billion time
  series}. In \bibinfo{booktitle}{\emph{2010 IEEE International Conference on
  Data Mining}}. IEEE, \bibinfo{pages}{58--67}.
\newblock


\bibitem[Camerra et~al\mbox{.}(2014)]%
        {isax2+}
\bibfield{author}{\bibinfo{person}{Alessandro Camerra}, \bibinfo{person}{Jin
  Shieh}, \bibinfo{person}{Themis Palpanas}, \bibinfo{person}{Thanawin
  Rakthanmanon}, {and} \bibinfo{person}{Eamonn Keogh}.}
  \bibinfo{year}{2014}\natexlab{}.
\newblock \showarticletitle{Beyond one billion time series: indexing and mining
  very large time series collections with $i$ sax2+}.
\newblock \bibinfo{journal}{\emph{Knowledge and information systems}}
  \bibinfo{volume}{39}, \bibinfo{number}{1} (\bibinfo{year}{2014}),
  \bibinfo{pages}{123--151}.
\newblock


\bibitem[Chatzakis et~al\mbox{.}(2023)]%
        {odyssey}
\bibfield{author}{\bibinfo{person}{Manos Chatzakis}, \bibinfo{person}{Panagiota
  Fatourou}, \bibinfo{person}{Eleftherios Kosmas}, \bibinfo{person}{Themis
  Palpanas}, {and} \bibinfo{person}{Botao Peng}.}
  \bibinfo{year}{2023}\natexlab{}.
\newblock \showarticletitle{{Odyssey: A Journey in the Land of Distributed Data
  Series Similarity Search}}.
\newblock \bibinfo{journal}{\emph{Proc. {VLDB} Endow.}} (\bibinfo{year}{2023}).
\newblock


\bibitem[Chen et~al\mbox{.}(2017)]%
        {nn-workshop}
\bibfield{author}{\bibinfo{person}{George Chen}, \bibinfo{person}{Christina
  Lee}, {and} \bibinfo{person}{Shah Devavrat}.}
  \bibinfo{year}{2017}\natexlab{}.
\newblock \showarticletitle{Nearest Neighbors for Modern Applications with
  Massive Data}. \bibinfo{howpublished}{\url{https://nn2017.mit.edu/}}. In
  \bibinfo{booktitle}{\emph{Proceedings of the 31rd International Conference on
  Neural Information Processing Systems Workshop}}.
\newblock


\bibitem[Chen and Shah(2018)]%
        {knn-prediction}
\bibfield{author}{\bibinfo{person}{George~H. Chen} {and}
  \bibinfo{person}{Devavrat Shah}.} \bibinfo{year}{2018}\natexlab{}.
\newblock \showarticletitle{Explaining the Success of Nearest Neighbor Methods
  in Prediction}.
\newblock \bibinfo{journal}{\emph{Foundations and Trends® in Machine
  Learning}} \bibinfo{volume}{10}, \bibinfo{number}{5-6}
  (\bibinfo{year}{2018}), \bibinfo{pages}{337--588}.
\newblock
\urldef\tempurl%
\url{https://doi.org/10.1561/2200000064}
\showDOI{\tempurl}


\bibitem[Chen et~al\mbox{.}(2021)]%
        {spann}
\bibfield{author}{\bibinfo{person}{Qi Chen}, \bibinfo{person}{Bing Zhao},
  \bibinfo{person}{Haidong Wang}, \bibinfo{person}{Mingqin Li},
  \bibinfo{person}{Chuanjie Liu}, \bibinfo{person}{Zhiyong Zheng},
  \bibinfo{person}{Mao Yang}, {and} \bibinfo{person}{Jingdong Wang}.}
  \bibinfo{year}{2021}\natexlab{}.
\newblock \showarticletitle{SPANN: Highly-efficient Billion-scale Approximate
  Nearest Neighborhood Search}.
\newblock \bibinfo{journal}{\emph{Advances in Neural Information Processing
  Systems}}  \bibinfo{volume}{34} (\bibinfo{year}{2021}).
\newblock


\bibitem[Ding et~al\mbox{.}(2008)]%
        {measure}
\bibfield{author}{\bibinfo{person}{Hui Ding}, \bibinfo{person}{Goce
  Trajcevski}, \bibinfo{person}{Peter Scheuermann}, \bibinfo{person}{Xiaoyue
  Wang}, {and} \bibinfo{person}{Eamonn Keogh}.}
  \bibinfo{year}{2008}\natexlab{}.
\newblock \showarticletitle{Querying and mining of time series data:
  experimental comparison of representations and distance measures}.
\newblock \bibinfo{journal}{\emph{Proceedings of the VLDB Endowment}}
  \bibinfo{volume}{1}, \bibinfo{number}{2} (\bibinfo{year}{2008}),
  \bibinfo{pages}{1542--1552}.
\newblock


\bibitem[Echihabi et~al\mbox{.}(2022)]%
        {DBLP:journals/pvldb/EchihabiFZPB22}
\bibfield{author}{\bibinfo{person}{Karima Echihabi}, \bibinfo{person}{Panagiota
  Fatourou}, \bibinfo{person}{Kostas Zoumpatianos}, \bibinfo{person}{Themis
  Palpanas}, {and} \bibinfo{person}{Houda Benbrahim}.}
  \bibinfo{year}{2022}\natexlab{}.
\newblock \showarticletitle{Hercules Against Data Series Similarity Search}.
\newblock \bibinfo{journal}{\emph{Proc. {VLDB} Endow.}} \bibinfo{volume}{15},
  \bibinfo{number}{10} (\bibinfo{year}{2022}), \bibinfo{pages}{2005--2018}.
\newblock


\bibitem[Echihabi et~al\mbox{.}(2023)]%
        {pros}
\bibfield{author}{\bibinfo{person}{Karima Echihabi},
  \bibinfo{person}{Theophanis Tsandilas}, \bibinfo{person}{Anna Gogolou},
  \bibinfo{person}{Anastasia Bezerianos}, {and} \bibinfo{person}{Themis
  Palpanas}.} \bibinfo{year}{2023}\natexlab{}.
\newblock \showarticletitle{{ProS: Data Series Progressive k-NN Similarity
  Search and Classification with Probabilistic Quality Guarantees}}.
\newblock \bibinfo{journal}{\emph{{VLDBJ}}} (\bibinfo{year}{2023}).
\newblock


\bibitem[Echihabi et~al\mbox{.}(2018)]%
        {hydra1}
\bibfield{author}{\bibinfo{person}{Karima Echihabi}, \bibinfo{person}{Kostas
  Zoumpatianos}, \bibinfo{person}{Themis Palpanas}, {and}
  \bibinfo{person}{Houda Benbrahim}.} \bibinfo{year}{2018}\natexlab{}.
\newblock \showarticletitle{The Lernaean Hydra of Data Series Similarity
  Search: An Experimental Evaluation of the State of the Art}.
\newblock \bibinfo{journal}{\emph{Proc. VLDB Endow.}} \bibinfo{volume}{12},
  \bibinfo{number}{2} (\bibinfo{date}{Oct.} \bibinfo{year}{2018}),
  \bibinfo{pages}{112–127}.
\newblock
\showISSN{2150-8097}


\bibitem[Echihabi et~al\mbox{.}(2019)]%
        {hydra2}
\bibfield{author}{\bibinfo{person}{Karima Echihabi}, \bibinfo{person}{Kostas
  Zoumpatianos}, \bibinfo{person}{Themis Palpanas}, {and}
  \bibinfo{person}{Houda Benbrahim}.} \bibinfo{year}{2019}\natexlab{}.
\newblock \showarticletitle{Return of the Lernaean Hydra: Experimental
  Evaluation of Data Series Approximate Similarity Search}.
\newblock \bibinfo{journal}{\emph{Proc. VLDB Endow.}} \bibinfo{volume}{13},
  \bibinfo{number}{3} (\bibinfo{date}{Nov.} \bibinfo{year}{2019}),
  \bibinfo{pages}{403–420}.
\newblock
\showISSN{2150-8097}


\bibitem[Fu et~al\mbox{.}(2021)]%
        {satellite}
\bibfield{author}{\bibinfo{person}{Cong Fu}, \bibinfo{person}{Changxu Wang},
  {and} \bibinfo{person}{Deng Cai}.} \bibinfo{year}{2021}\natexlab{}.
\newblock \showarticletitle{High Dimensional Similarity Search with Satellite
  System Graph: Efficiency, Scalability, and Unindexed Query Compatibility}.
\newblock \bibinfo{journal}{\emph{IEEE Transactions on Pattern Analysis and
  Machine Intelligence}} (\bibinfo{year}{2021}), \bibinfo{pages}{1--1}.
\newblock


\bibitem[Fu et~al\mbox{.}(2019)]%
        {nsg}
\bibfield{author}{\bibinfo{person}{Cong Fu}, \bibinfo{person}{Chao Xiang},
  \bibinfo{person}{Changxu Wang}, {and} \bibinfo{person}{Deng Cai}.}
  \bibinfo{year}{2019}\natexlab{}.
\newblock \showarticletitle{Fast Approximate Nearest Neighbor Search with the
  Navigating Spreading-out Graph}.
\newblock \bibinfo{journal}{\emph{Proc. VLDB Endow.}} \bibinfo{volume}{12},
  \bibinfo{number}{5} (\bibinfo{date}{Jan.} \bibinfo{year}{2019}),
  \bibinfo{pages}{461–474}.
\newblock
\showISSN{2150-8097}


\bibitem[Ge et~al\mbox{.}(2013)]%
        {opt-pq}
\bibfield{author}{\bibinfo{person}{Tiezheng Ge}, \bibinfo{person}{Kaiming He},
  \bibinfo{person}{Qifa Ke}, {and} \bibinfo{person}{Jian Sun}.}
  \bibinfo{year}{2013}\natexlab{}.
\newblock \showarticletitle{Optimized product quantization}.
\newblock \bibinfo{journal}{\emph{IEEE transactions on pattern analysis and
  machine intelligence}} \bibinfo{volume}{36}, \bibinfo{number}{4}
  (\bibinfo{year}{2013}), \bibinfo{pages}{744--755}.
\newblock


\bibitem[Gong et~al\mbox{.}(2020)]%
        {idec}
\bibfield{author}{\bibinfo{person}{Long Gong}, \bibinfo{person}{Huayi Wang},
  \bibinfo{person}{Mitsunori Ogihara}, {and} \bibinfo{person}{Jun Xu}.}
  \bibinfo{year}{2020}\natexlab{}.
\newblock \showarticletitle{iDEC: indexable distance estimating codes for
  approximate nearest neighbor search}.
\newblock \bibinfo{journal}{\emph{Proceedings of the VLDB Endowment}}
  \bibinfo{volume}{13}, \bibinfo{number}{9} (\bibinfo{year}{2020}).
\newblock


\bibitem[Huang et~al\mbox{.}(2015)]%
        {qa-lsh}
\bibfield{author}{\bibinfo{person}{Qiang Huang}, \bibinfo{person}{Jianlin
  Feng}, \bibinfo{person}{Yikai Zhang}, \bibinfo{person}{Qiong Fang}, {and}
  \bibinfo{person}{Wilfred Ng}.} \bibinfo{year}{2015}\natexlab{}.
\newblock \showarticletitle{Query-aware locality-sensitive hashing for
  approximate nearest neighbor search}.
\newblock \bibinfo{journal}{\emph{Proceedings of the VLDB Endowment}}
  \bibinfo{volume}{9}, \bibinfo{number}{1} (\bibinfo{year}{2015}),
  \bibinfo{pages}{1--12}.
\newblock


\bibitem[Jegou et~al\mbox{.}(2010)]%
        {pq}
\bibfield{author}{\bibinfo{person}{Herve Jegou}, \bibinfo{person}{Matthijs
  Douze}, {and} \bibinfo{person}{Cordelia Schmid}.}
  \bibinfo{year}{2010}\natexlab{}.
\newblock \showarticletitle{Product quantization for nearest neighbor search}.
\newblock \bibinfo{journal}{\emph{IEEE transactions on pattern analysis and
  machine intelligence}} \bibinfo{volume}{33}, \bibinfo{number}{1}
  (\bibinfo{year}{2010}), \bibinfo{pages}{117--128}.
\newblock


\bibitem[Jo et~al\mbox{.}(2020)]%
        {DBLP:journals/tvcg/JoSF20}
\bibfield{author}{\bibinfo{person}{Jaemin Jo}, \bibinfo{person}{Jinwook Seo},
  {and} \bibinfo{person}{Jean{-}Daniel Fekete}.}
  \bibinfo{year}{2020}\natexlab{}.
\newblock \showarticletitle{{PANENE:} {A} Progressive Algorithm for Indexing
  and Querying Approximate k-Nearest Neighbors}.
\newblock \bibinfo{journal}{\emph{{IEEE} Trans. Vis. Comput. Graph.}}
  \bibinfo{volume}{26}, \bibinfo{number}{2} (\bibinfo{year}{2020}),
  \bibinfo{pages}{1347--1360}.
\newblock


\bibitem[Johnson et~al\mbox{.}(2016)]%
        {ecg}
\bibfield{author}{\bibinfo{person}{Alistair~EW Johnson}, \bibinfo{person}{Tom~J
  Pollard}, \bibinfo{person}{Lu Shen}, \bibinfo{person}{Li-wei~H Lehman},
  \bibinfo{person}{Mengling Feng}, \bibinfo{person}{Mohammad Ghassemi},
  \bibinfo{person}{Benjamin Moody}, \bibinfo{person}{Peter Szolovits},
  \bibinfo{person}{Leo Anthony~Celi}, {and} \bibinfo{person}{Roger~G Mark}.}
  \bibinfo{year}{2016}\natexlab{}.
\newblock \showarticletitle{MIMIC-III, a freely accessible critical care
  database}.
\newblock \bibinfo{journal}{\emph{Scientific data}} \bibinfo{volume}{3},
  \bibinfo{number}{1} (\bibinfo{year}{2016}), \bibinfo{pages}{1--9}.
\newblock


\bibitem[Keogh(2006)]%
        {talking3}
\bibfield{author}{\bibinfo{person}{Eamonn Keogh}.}
  \bibinfo{year}{2006}\natexlab{}.
\newblock \showarticletitle{A decade of progress in indexing and mining large
  time series databases}. In \bibinfo{booktitle}{\emph{Proceedings of the 32nd
  international conference on Very large data bases}}.
  \bibinfo{pages}{1268--1268}.
\newblock


\bibitem[Keogh et~al\mbox{.}(2001)]%
        {paa}
\bibfield{author}{\bibinfo{person}{Eamonn Keogh}, \bibinfo{person}{Kaushik
  Chakrabarti}, \bibinfo{person}{Michael Pazzani}, {and}
  \bibinfo{person}{Sharad Mehrotra}.} \bibinfo{year}{2001}\natexlab{}.
\newblock \showarticletitle{Dimensionality reduction for fast similarity search
  in large time series databases}.
\newblock \bibinfo{journal}{\emph{Knowledge and information Systems}}
  \bibinfo{volume}{3}, \bibinfo{number}{3} (\bibinfo{year}{2001}),
  \bibinfo{pages}{263--286}.
\newblock


\bibitem[Kondylakis et~al\mbox{.}(2018)]%
        {coconut}
\bibfield{author}{\bibinfo{person}{Haridimos Kondylakis}, \bibinfo{person}{Niv
  Dayan}, \bibinfo{person}{Kostas Zoumpatianos}, {and} \bibinfo{person}{Themis
  Palpanas}.} \bibinfo{year}{2018}\natexlab{}.
\newblock \showarticletitle{Coconut: A Scalable Bottom-up Approach for Building
  Data Series Indexes}.
\newblock \bibinfo{journal}{\emph{Proc. VLDB Endow.}} \bibinfo{volume}{11},
  \bibinfo{number}{6} (\bibinfo{date}{Feb.} \bibinfo{year}{2018}),
  \bibinfo{pages}{677–690}.
\newblock
\showISSN{2150-8097}


\bibitem[Kondylakis et~al\mbox{.}(2019)]%
        {DBLP:journals/vldb/KondylakisDZP19}
\bibfield{author}{\bibinfo{person}{Haridimos Kondylakis}, \bibinfo{person}{Niv
  Dayan}, \bibinfo{person}{Kostas Zoumpatianos}, {and} \bibinfo{person}{Themis
  Palpanas}.} \bibinfo{year}{2019}\natexlab{}.
\newblock \showarticletitle{Coconut: sortable summarizations for scalable
  indexes over static and streaming data series}.
\newblock \bibinfo{journal}{\emph{{VLDB} J.}} \bibinfo{volume}{28},
  \bibinfo{number}{6} (\bibinfo{year}{2019}), \bibinfo{pages}{847--869}.
\newblock
\urldef\tempurl%
\url{https://doi.org/10.1007/s00778-019-00573-w}
\showDOI{\tempurl}


\bibitem[Korn et~al\mbox{.}(2001)]%
        {DBLP:journals/tkde/KornPF01}
\bibfield{author}{\bibinfo{person}{Flip Korn}, \bibinfo{person}{Bernd{-}Uwe
  Pagel}, {and} \bibinfo{person}{Christos Faloutsos}.}
  \bibinfo{year}{2001}\natexlab{}.
\newblock \showarticletitle{On the 'Dimensionality Curse' and the
  'Self-Similarity Blessing'}.
\newblock \bibinfo{journal}{\emph{{IEEE} Trans. Knowl. Data Eng.}}
  \bibinfo{volume}{13}, \bibinfo{number}{1} (\bibinfo{year}{2001}),
  \bibinfo{pages}{96--111}.
\newblock
\urldef\tempurl%
\url{https://doi.org/10.1109/69.908983}
\showDOI{\tempurl}


\bibitem[Levchenko et~al\mbox{.}(2018)]%
        {parsketch}
\bibfield{author}{\bibinfo{person}{Oleksandra Levchenko},
  \bibinfo{person}{Djamel-Edine Yagoubi}, \bibinfo{person}{Reza Akbarinia},
  \bibinfo{person}{Florent Masseglia}, \bibinfo{person}{Boyan Kolev}, {and}
  \bibinfo{person}{Dennis Shasha}.} \bibinfo{year}{2018}\natexlab{}.
\newblock \showarticletitle{Spark-parsketch: a massively distributed indexing
  of time series datasets}. In \bibinfo{booktitle}{\emph{Proceedings of the
  27th ACM International Conference on Information and Knowledge Management}}.
  \bibinfo{pages}{1951--1954}.
\newblock


\bibitem[Li et~al\mbox{.}(2019)]%
        {li-evaluation}
\bibfield{author}{\bibinfo{person}{Wen Li}, \bibinfo{person}{Ying Zhang},
  \bibinfo{person}{Yifang Sun}, \bibinfo{person}{Wei Wang},
  \bibinfo{person}{Mingjie Li}, \bibinfo{person}{Wenjie Zhang}, {and}
  \bibinfo{person}{Xuemin Lin}.} \bibinfo{year}{2019}\natexlab{}.
\newblock \showarticletitle{Approximate nearest neighbor search on high
  dimensional data—experiments, analyses, and improvement}.
\newblock \bibinfo{journal}{\emph{IEEE Transactions on Knowledge and Data
  Engineering}} \bibinfo{volume}{32}, \bibinfo{number}{8}
  (\bibinfo{year}{2019}), \bibinfo{pages}{1475--1488}.
\newblock


\bibitem[Linardi and Palpanas(2018)]%
        {ulisse}
\bibfield{author}{\bibinfo{person}{Michele Linardi} {and}
  \bibinfo{person}{Themis Palpanas}.} \bibinfo{year}{2018}\natexlab{}.
\newblock \showarticletitle{Scalable, Variable-Length Similarity Search in Data
  Series: The ULISSE Approach}.
\newblock \bibinfo{journal}{\emph{Proc. VLDB Endow.}} \bibinfo{volume}{11},
  \bibinfo{number}{13} (\bibinfo{year}{2018}), \bibinfo{pages}{2236–2248}.
\newblock
\urldef\tempurl%
\url{https://doi.org/10.14778/3275366.3284968}
\showDOI{\tempurl}


\bibitem[Linardi and Palpanas(2020)]%
        {DBLP:journals/vldb/LinardiP20}
\bibfield{author}{\bibinfo{person}{Michele Linardi} {and}
  \bibinfo{person}{Themis Palpanas}.} \bibinfo{year}{2020}\natexlab{}.
\newblock \showarticletitle{Scalable data series subsequence matching with
  {ULISSE}}.
\newblock \bibinfo{journal}{\emph{{VLDB} J.}} \bibinfo{volume}{29},
  \bibinfo{number}{6} (\bibinfo{year}{2020}), \bibinfo{pages}{1449--1474}.
\newblock


\bibitem[Liu et~al\mbox{.}(2021)]%
        {eilsh}
\bibfield{author}{\bibinfo{person}{Wanqi Liu}, \bibinfo{person}{Hanchen Wang},
  \bibinfo{person}{Ying Zhang}, \bibinfo{person}{Wei Wang}, \bibinfo{person}{Lu
  Qin}, {and} \bibinfo{person}{Xuemin Lin}.} \bibinfo{year}{2021}\natexlab{}.
\newblock \showarticletitle{EI-LSH: An early-termination driven I/O efficient
  incremental c-approximate nearest neighbor search}.
\newblock \bibinfo{journal}{\emph{The VLDB Journal}} \bibinfo{volume}{30},
  \bibinfo{number}{2} (\bibinfo{year}{2021}), \bibinfo{pages}{215--235}.
\newblock


\bibitem[Malkov et~al\mbox{.}(2014)]%
        {nsw}
\bibfield{author}{\bibinfo{person}{Yury Malkov}, \bibinfo{person}{Alexander
  Ponomarenko}, \bibinfo{person}{Andrey Logvinov}, {and}
  \bibinfo{person}{Vladimir Krylov}.} \bibinfo{year}{2014}\natexlab{}.
\newblock \showarticletitle{Approximate nearest neighbor algorithm based on
  navigable small world graphs}.
\newblock \bibinfo{journal}{\emph{Information Systems}}  \bibinfo{volume}{45}
  (\bibinfo{year}{2014}), \bibinfo{pages}{61--68}.
\newblock


\bibitem[Malkov and Yashunin(2018)]%
        {hnsw}
\bibfield{author}{\bibinfo{person}{Yu~A Malkov} {and} \bibinfo{person}{Dmitry~A
  Yashunin}.} \bibinfo{year}{2018}\natexlab{}.
\newblock \showarticletitle{Efficient and robust approximate nearest neighbor
  search using hierarchical navigable small world graphs}.
\newblock \bibinfo{journal}{\emph{IEEE transactions on pattern analysis and
  machine intelligence}} \bibinfo{volume}{42}, \bibinfo{number}{4}
  (\bibinfo{year}{2018}), \bibinfo{pages}{824--836}.
\newblock


\bibitem[Palpanas(2015)]%
        {talking2}
\bibfield{author}{\bibinfo{person}{Themis Palpanas}.}
  \bibinfo{year}{2015}\natexlab{}.
\newblock \showarticletitle{Data series management: The road to big sequence
  analytics}.
\newblock \bibinfo{journal}{\emph{ACM SIGMOD Record}} \bibinfo{volume}{44},
  \bibinfo{number}{2} (\bibinfo{year}{2015}), \bibinfo{pages}{47--52}.
\newblock


\bibitem[Palpanas(2016)]%
        {DBLP:conf/sofsem/Palpanas16}
\bibfield{author}{\bibinfo{person}{Themis Palpanas}.}
  \bibinfo{year}{2016}\natexlab{}.
\newblock \showarticletitle{Big Sequence Management: {A} glimpse of the Past,
  the Present, and the Future}. In \bibinfo{booktitle}{\emph{{SOFSEM} 2016:
  Theory and Practice of Computer Science - 42nd International Conference on
  Current Trends in Theory and Practice of Computer Science}}
  \emph{(\bibinfo{series}{Lecture Notes in Computer Science},
  Vol.~\bibinfo{volume}{9587})}. \bibinfo{pages}{63--80}.
\newblock


\bibitem[Palpanas(2020)]%
        {c19-isip-Palpanas-isaxfamily}
\bibfield{author}{\bibinfo{person}{Themis Palpanas}.}
  \bibinfo{year}{2020}\natexlab{}.
\newblock \showarticletitle{Evolution of a Data Series Index}. In
  \bibinfo{booktitle}{\emph{Information Search, Integration, and
  Personalization}}. \bibinfo{publisher}{Springer International Publishing},
  \bibinfo{address}{Cham}, \bibinfo{pages}{68--83}.
\newblock


\bibitem[Palpanas and Beckmann(2019)]%
        {DBLP:journals/sigmod/PalpanasB19}
\bibfield{author}{\bibinfo{person}{Themis Palpanas} {and}
  \bibinfo{person}{Volker Beckmann}.} \bibinfo{year}{2019}\natexlab{}.
\newblock \showarticletitle{Report on the First and Second Interdisciplinary
  Time Series Analysis Workshop {(ITISA)}}.
\newblock \bibinfo{journal}{\emph{{SIGMOD} Rec.}} \bibinfo{volume}{48},
  \bibinfo{number}{3} (\bibinfo{year}{2019}), \bibinfo{pages}{36--40}.
\newblock
\urldef\tempurl%
\url{https://doi.org/10.1145/3377391.3377400}
\showDOI{\tempurl}


\bibitem[Paparrizos et~al\mbox{.}(2022)]%
        {new-pq}
\bibfield{author}{\bibinfo{person}{John Paparrizos}, \bibinfo{person}{Ikraduya
  Edian}, \bibinfo{person}{Chunwei Liu}, \bibinfo{person}{Aaron~J. Elmore},
  {and} \bibinfo{person}{Michael~J. Franklin}.}
  \bibinfo{year}{2022}\natexlab{}.
\newblock \showarticletitle{Fast Adaptive Similarity Search through
  Variance-Aware Quantization}. In \bibinfo{booktitle}{\emph{2022 IEEE 38th
  International Conference on Data Engineering (ICDE)}}. IEEE.
\newblock


\bibitem[Peng et~al\mbox{.}(2018)]%
        {paris}
\bibfield{author}{\bibinfo{person}{Botao Peng}, \bibinfo{person}{Panagiota
  Fatourou}, {and} \bibinfo{person}{Themis Palpanas}.}
  \bibinfo{year}{2018}\natexlab{}.
\newblock \showarticletitle{Paris: The next destination for fast data series
  indexing and query answering}. In \bibinfo{booktitle}{\emph{2018 IEEE
  International Conference on Big Data (Big Data)}}. IEEE,
  \bibinfo{pages}{791--800}.
\newblock


\bibitem[Peng et~al\mbox{.}(2020a)]%
        {messi}
\bibfield{author}{\bibinfo{person}{Botao Peng}, \bibinfo{person}{Panagiota
  Fatourou}, {and} \bibinfo{person}{Themis Palpanas}.}
  \bibinfo{year}{2020}\natexlab{a}.
\newblock \showarticletitle{Messi: In-memory data series indexing}. In
  \bibinfo{booktitle}{\emph{2020 IEEE 36th International Conference on Data
  Engineering (ICDE)}}. IEEE, \bibinfo{pages}{337--348}.
\newblock


\bibitem[Peng et~al\mbox{.}(2020b)]%
        {paris+}
\bibfield{author}{\bibinfo{person}{Botao Peng}, \bibinfo{person}{Panagiota
  Fatourou}, {and} \bibinfo{person}{Themis Palpanas}.}
  \bibinfo{year}{2020}\natexlab{b}.
\newblock \showarticletitle{Paris+: Data series indexing on multi-core
  architectures}.
\newblock \bibinfo{journal}{\emph{IEEE Transactions on Knowledge and Data
  Engineering}} \bibinfo{volume}{33}, \bibinfo{number}{5}
  (\bibinfo{year}{2020}), \bibinfo{pages}{2151--2164}.
\newblock


\bibitem[Peng et~al\mbox{.}(2021a)]%
        {DBLP:journals/vldb/PengFP21}
\bibfield{author}{\bibinfo{person}{Botao Peng}, \bibinfo{person}{Panagiota
  Fatourou}, {and} \bibinfo{person}{Themis Palpanas}.}
  \bibinfo{year}{2021}\natexlab{a}.
\newblock \showarticletitle{Fast data series indexing for in-memory data}.
\newblock \bibinfo{journal}{\emph{{VLDB} J.}} \bibinfo{volume}{30},
  \bibinfo{number}{6} (\bibinfo{year}{2021}), \bibinfo{pages}{1041--1067}.
\newblock


\bibitem[Peng et~al\mbox{.}(2021b)]%
        {sing}
\bibfield{author}{\bibinfo{person}{Botao Peng}, \bibinfo{person}{Panagiota
  Fatourou}, {and} \bibinfo{person}{Themis Palpanas}.}
  \bibinfo{year}{2021}\natexlab{b}.
\newblock \showarticletitle{SING: Sequence Indexing Using GPUs}. In
  \bibinfo{booktitle}{\emph{2021 IEEE 37th International Conference on Data
  Engineering (ICDE)}}. IEEE, \bibinfo{pages}{1883--1888}.
\newblock


\bibitem[Rakthanmanon et~al\mbox{.}(2012)]%
        {dtw}
\bibfield{author}{\bibinfo{person}{Thanawin Rakthanmanon},
  \bibinfo{person}{Bilson Campana}, \bibinfo{person}{Abdullah Mueen},
  \bibinfo{person}{Gustavo Batista}, \bibinfo{person}{Brandon Westover},
  \bibinfo{person}{Qiang Zhu}, \bibinfo{person}{Jesin Zakaria}, {and}
  \bibinfo{person}{Eamonn Keogh}.} \bibinfo{year}{2012}\natexlab{}.
\newblock \showarticletitle{Searching and Mining Trillions of Time Series
  Subsequences under Dynamic Time Warping}. In
  \bibinfo{booktitle}{\emph{Proceedings of the 18th ACM SIGKDD International
  Conference on Knowledge Discovery and Data Mining}}
  \emph{(\bibinfo{series}{KDD '12})}. \bibinfo{pages}{262–270}.
\newblock
\urldef\tempurl%
\url{https://doi.org/10.1145/2339530.2339576}
\showDOI{\tempurl}


\bibitem[Schubert et~al\mbox{.}(2015)]%
        {knn-outlier}
\bibfield{author}{\bibinfo{person}{Erich Schubert}, \bibinfo{person}{Arthur
  Zimek}, {and} \bibinfo{person}{Hans-Peter Kriegel}.}
  \bibinfo{year}{2015}\natexlab{}.
\newblock \showarticletitle{Fast and Scalable Outlier Detection with
  Approximate Nearest Neighbor Ensembles}. In
  \bibinfo{booktitle}{\emph{Database Systems for Advanced Applications}}.
  \bibinfo{pages}{19--36}.
\newblock


\bibitem[Shannon(1948)]%
        {entropy}
\bibfield{author}{\bibinfo{person}{Claude~Elwood Shannon}.}
  \bibinfo{year}{1948}\natexlab{}.
\newblock \showarticletitle{A mathematical theory of communication}.
\newblock \bibinfo{journal}{\emph{The Bell system technical journal}}
  \bibinfo{volume}{27}, \bibinfo{number}{3} (\bibinfo{year}{1948}),
  \bibinfo{pages}{379--423}.
\newblock


\bibitem[Shieh and Keogh(2008)]%
        {isax}
\bibfield{author}{\bibinfo{person}{Jin Shieh} {and} \bibinfo{person}{Eamonn
  Keogh}.} \bibinfo{year}{2008}\natexlab{}.
\newblock \showarticletitle{iSAX: indexing and mining terabyte sized time
  series}. In \bibinfo{booktitle}{\emph{Proceedings of the 14th ACM SIGKDD
  international conference on Knowledge discovery and data mining}}.
  \bibinfo{pages}{623--631}.
\newblock


\bibitem[Subramanya et~al\mbox{.}(2019)]%
        {diskann}
\bibfield{author}{\bibinfo{person}{Suhas~Jayaram Subramanya},
  \bibinfo{person}{Rohan Kadekodi}, \bibinfo{person}{Ravishankar Krishaswamy},
  {and} \bibinfo{person}{Harsha~Vardhan Simhadri}.}
  \bibinfo{year}{2019}\natexlab{}.
\newblock \showarticletitle{Diskann: Fast accurate billion-point nearest
  neighbor search on a single node}. In \bibinfo{booktitle}{\emph{Proceedings
  of the 33rd International Conference on Neural Information Processing
  Systems}}. \bibinfo{pages}{13766--13776}.
\newblock


\bibitem[Sun et~al\mbox{.}(2014)]%
        {srs}
\bibfield{author}{\bibinfo{person}{Yifang Sun}, \bibinfo{person}{Wei Wang},
  \bibinfo{person}{Jianbin Qin}, \bibinfo{person}{Ying Zhang}, {and}
  \bibinfo{person}{Xuemin Lin}.} \bibinfo{year}{2014}\natexlab{}.
\newblock \showarticletitle{SRS: Solving $c$-Approximate Nearest Neighbor
  Queries in High Dimensional Euclidean Space with a Tiny Index}.
\newblock \bibinfo{journal}{\emph{Proc. VLDB Endow.}} \bibinfo{volume}{8},
  \bibinfo{number}{1} (\bibinfo{date}{Sept.} \bibinfo{year}{2014}),
  \bibinfo{pages}{1–12}.
\newblock
\showISSN{2150-8097}


\bibitem[Tan et~al\mbox{.}(2017)]%
        {app1}
\bibfield{author}{\bibinfo{person}{Chang~Wei Tan}, \bibinfo{person}{Geoffrey~I
  Webb}, {and} \bibinfo{person}{Fran{\c{c}}ois Petitjean}.}
  \bibinfo{year}{2017}\natexlab{}.
\newblock \showarticletitle{Indexing and classifying gigabytes of time series
  under time warping}. In \bibinfo{booktitle}{\emph{Proceedings of the 2017
  SIAM international conference on data mining}}. SIAM,
  \bibinfo{pages}{282--290}.
\newblock


\bibitem[Turpin and Scholer(2006)]%
        {map}
\bibfield{author}{\bibinfo{person}{Andrew Turpin} {and} \bibinfo{person}{Falk
  Scholer}.} \bibinfo{year}{2006}\natexlab{}.
\newblock \showarticletitle{User performance versus precision measures for
  simple search tasks}. In \bibinfo{booktitle}{\emph{Proceedings of the 29th
  annual international ACM SIGIR conference on Research and development in
  information retrieval}}. \bibinfo{pages}{11--18}.
\newblock


\bibitem[Vision({[n.\,d.]})]%
        {deep}
\bibfield{author}{\bibinfo{person}{Skoltech~Computer Vision}.}
  \bibinfo{year}{[n.\,d.]}\natexlab{}.
\newblock \bibinfo{title}{Deep billion-scale indexing}.
\newblock
  \bibinfo{howpublished}{\url{http://sites.skoltech.ru/compvision/noimi}
  Accessed March 14, 2022}.
\newblock


\bibitem[Wang et~al\mbox{.}(2021)]%
        {wang-evaluation}
\bibfield{author}{\bibinfo{person}{Mengzhao Wang}, \bibinfo{person}{Xiaoliang
  Xu}, \bibinfo{person}{Qiang Yue}, {and} \bibinfo{person}{Yuxiang Wang}.}
  \bibinfo{year}{2021}\natexlab{}.
\newblock \showarticletitle{A Comprehensive Survey and Experimental Comparison
  of Graph-Based Approximate Nearest Neighbor Search}.
\newblock \bibinfo{journal}{\emph{arXiv preprint arXiv:2101.12631}}
  (\bibinfo{year}{2021}).
\newblock


\bibitem[Wang and Palpanas(2021)]%
        {qt}
\bibfield{author}{\bibinfo{person}{Qitong Wang} {and} \bibinfo{person}{Themis
  Palpanas}.} \bibinfo{year}{2021}\natexlab{}.
\newblock \showarticletitle{Deep Learning Embeddings for Data Series Similarity
  Search} \emph{(\bibinfo{series}{KDD '21})}. \bibinfo{publisher}{ACM},
  \bibinfo{address}{NY, USA}, \bibinfo{pages}{1708–1716}.
\newblock
\showISBNx{9781450383325}
\urldef\tempurl%
\url{https://doi.org/10.1145/3447548.3467317}
\showURL{%
\tempurl}


\bibitem[Wang et~al\mbox{.}(2022)]%
        {DBLP:journals/pvldb/0003WNP22}
\bibfield{author}{\bibinfo{person}{Qitong Wang}, \bibinfo{person}{Stephen
  Whitmarsh}, \bibinfo{person}{Vincent Navarro}, {and} \bibinfo{person}{Themis
  Palpanas}.} \bibinfo{year}{2022}\natexlab{}.
\newblock \showarticletitle{iEDeaL: {A} Deep Learning Framework for Detecting
  Highly Imbalanced Interictal Epileptiform Discharges}.
\newblock \bibinfo{journal}{\emph{Proc. {VLDB} Endow.}} \bibinfo{volume}{16},
  \bibinfo{number}{3} (\bibinfo{year}{2022}), \bibinfo{pages}{480--490}.
\newblock


\bibitem[Wang et~al\mbox{.}(2013)]%
        {ds-tree}
\bibfield{author}{\bibinfo{person}{Yang Wang}, \bibinfo{person}{Peng Wang},
  \bibinfo{person}{Jian Pei}, \bibinfo{person}{Wei Wang}, {and}
  \bibinfo{person}{Sheng Huang}.} \bibinfo{year}{2013}\natexlab{}.
\newblock \showarticletitle{A data-adaptive and dynamic segmentation index for
  whole matching on time series}.
\newblock \bibinfo{journal}{\emph{Proceedings of the VLDB Endowment}}
  \bibinfo{volume}{6}, \bibinfo{number}{10} (\bibinfo{year}{2013}),
  \bibinfo{pages}{793--804}.
\newblock


\bibitem[Yagoubi et~al\mbox{.}(2017)]%
        {dpisax}
\bibfield{author}{\bibinfo{person}{Djamel~Edine Yagoubi}, \bibinfo{person}{Reza
  Akbarinia}, \bibinfo{person}{Florent Masseglia}, {and}
  \bibinfo{person}{Themis Palpanas}.} \bibinfo{year}{2017}\natexlab{}.
\newblock \showarticletitle{Dpisax: Massively distributed partitioned isax}. In
  \bibinfo{booktitle}{\emph{2017 IEEE International Conference on Data Mining
  (ICDM)}}. IEEE, \bibinfo{pages}{1135--1140}.
\newblock


\bibitem[Yagoubi et~al\mbox{.}(2020)]%
        {DBLP:journals/tkde/YagoubiAMP20}
\bibfield{author}{\bibinfo{person}{Djamel~Edine Yagoubi}, \bibinfo{person}{Reza
  Akbarinia}, \bibinfo{person}{Florent Masseglia}, {and}
  \bibinfo{person}{Themis Palpanas}.} \bibinfo{year}{2020}\natexlab{}.
\newblock \showarticletitle{Massively Distributed Time Series Indexing and
  Querying}.
\newblock \bibinfo{journal}{\emph{{IEEE} Trans. Knowl. Data Eng.}}
  \bibinfo{volume}{32}, \bibinfo{number}{1} (\bibinfo{year}{2020}),
  \bibinfo{pages}{108--120}.
\newblock
\urldef\tempurl%
\url{https://doi.org/10.1109/TKDE.2018.2880215}
\showDOI{\tempurl}


\bibitem[Zhang et~al\mbox{.}(2019)]%
        {tardis}
\bibfield{author}{\bibinfo{person}{Liang Zhang}, \bibinfo{person}{Noura
  Alghamdi}, \bibinfo{person}{Mohamed~Y Eltabakh}, {and}
  \bibinfo{person}{Elke~A Rundensteiner}.} \bibinfo{year}{2019}\natexlab{}.
\newblock \showarticletitle{TARDIS: Distributed indexing framework for big time
  series data}. In \bibinfo{booktitle}{\emph{2019 IEEE 35th International
  Conference on Data Engineering (ICDE)}}. IEEE, \bibinfo{pages}{1202--1213}.
\newblock


\bibitem[Zhao et~al\mbox{.}(2021)]%
        {knn-softmax}
\bibfield{author}{\bibinfo{person}{Kang Zhao}, \bibinfo{person}{Liuyihan Song},
  \bibinfo{person}{Yingya Zhang}, \bibinfo{person}{Pan Pan},
  \bibinfo{person}{Yinghui Xu}, {and} \bibinfo{person}{Rong Jin}.}
  \bibinfo{year}{2021}\natexlab{}.
\newblock \showarticletitle{ANN Softmax: Acceleration of Extreme Classification
  Training}.
\newblock \bibinfo{journal}{\emph{Proc. VLDB Endow.}} \bibinfo{volume}{15},
  \bibinfo{number}{1} (\bibinfo{year}{2021}), \bibinfo{pages}{1–10}.
\newblock
\urldef\tempurl%
\url{https://doi.org/10.14778/3485450.3485451}
\showDOI{\tempurl}


\bibitem[Zoumpatianos et~al\mbox{.}(2014)]%
        {ads}
\bibfield{author}{\bibinfo{person}{Kostas Zoumpatianos},
  \bibinfo{person}{Stratos Idreos}, {and} \bibinfo{person}{Themis Palpanas}.}
  \bibinfo{year}{2014}\natexlab{}.
\newblock \showarticletitle{Indexing for interactive exploration of big data
  series}. In \bibinfo{booktitle}{\emph{Proceedings of the 2014 ACM SIGMOD
  international conference on Management of data}}.
  \bibinfo{pages}{1555--1566}.
\newblock


\bibitem[Zoumpatianos et~al\mbox{.}(2016)]%
        {ads-full}
\bibfield{author}{\bibinfo{person}{Kostas Zoumpatianos},
  \bibinfo{person}{Stratos Idreos}, {and} \bibinfo{person}{Themis Palpanas}.}
  \bibinfo{year}{2016}\natexlab{}.
\newblock \showarticletitle{ADS: the adaptive data series index}.
\newblock \bibinfo{journal}{\emph{The VLDB Journal}} \bibinfo{volume}{25},
  \bibinfo{number}{6} (\bibinfo{year}{2016}), \bibinfo{pages}{843--866}.
\newblock


\bibitem[Zoumpatianos et~al\mbox{.}(2018)]%
        {DBLP:journals/vldb/ZoumpatianosLIP18}
\bibfield{author}{\bibinfo{person}{Kostas Zoumpatianos}, \bibinfo{person}{Yin
  Lou}, \bibinfo{person}{Ioana Ileana}, \bibinfo{person}{Themis Palpanas},
  {and} \bibinfo{person}{Johannes Gehrke}.} \bibinfo{year}{2018}\natexlab{}.
\newblock \showarticletitle{Generating data series query workloads}.
\newblock \bibinfo{journal}{\emph{{VLDB} J.}} \bibinfo{volume}{27},
  \bibinfo{number}{6} (\bibinfo{year}{2018}), \bibinfo{pages}{823--846}.
\newblock
\urldef\tempurl%
\url{https://doi.org/10.1007/s00778-018-0513-x}
\showDOI{\tempurl}


\bibitem[Zoumpatianos et~al\mbox{.}(2015)]%
        {hardness}
\bibfield{author}{\bibinfo{person}{Kostas Zoumpatianos}, \bibinfo{person}{Yin
  Lou}, \bibinfo{person}{Themis Palpanas}, {and} \bibinfo{person}{Johannes
  Gehrke}.} \bibinfo{year}{2015}\natexlab{}.
\newblock \showarticletitle{Query workloads for data series indexes}. In
  \bibinfo{booktitle}{\emph{Proceedings of the 21th ACM SIGKDD International
  Conference on Knowledge Discovery and Data Mining}}.
  \bibinfo{pages}{1603--1612}.
\newblock


\bibitem[Zoumpatianos and Palpanas(2018)]%
        {talking1}
\bibfield{author}{\bibinfo{person}{Kostas Zoumpatianos} {and}
  \bibinfo{person}{Themis Palpanas}.} \bibinfo{year}{2018}\natexlab{}.
\newblock \showarticletitle{Data series management: Fulfilling the need for big
  sequence analytics}. In \bibinfo{booktitle}{\emph{2018 IEEE 34th
  International Conference on Data Engineering (ICDE)}}. IEEE,
  \bibinfo{pages}{1677--1678}.
\newblock


\end{thebibliography}
\end{document}